\documentclass[aps,prd,twoside,twocolumn,superscriptaddress,floatfix,nofootinbib]{revtex4}
\usepackage{amsmath}
\usepackage{bm}
\usepackage{xcolor}
\usepackage{sidecap}
\usepackage{hyperref}
\usepackage{comment}

\newcommand\beq{\begin{equation}}
\newcommand\eeq{\end{equation}}
\newcommand\beqn{\begin{eqnarray}}
\newcommand\eeqn{\end{eqnarray}}

\newcommand\nn{\nonumber}

\newcommand{\ba}{\begin{eqnarray}}
\newcommand{\ea}{\end{eqnarray}}
\newcommand{\be}{\begin{equation}}
\newcommand{\ee}{\end{equation}}

\newcommand{\barr}{\begin{array}}
\newcommand{\earr}{\end{array}}

\newcommand\lsim{\mathrel{\rlap{\lower4pt\hbox{\hskip1pt$\sim$}}
        \raise1pt\hbox{$<$}}}
\newcommand\gsim{\mathrel{\rlap{\lower4pt\hbox{\hskip1pt$\sim$}}
        \raise1pt\hbox{$>$}}}

\newcommand{\jcap}{{J.~Cosm.~Astrop.~Phys.}}

\newcommand{\aap}{{Astron.~Astrophys.}}
\newcommand{\apjl}{{Astrophys.~J.~Lett.}}
\newcommand{\apjs}{{Astrophys.~J.~Supp.}}

\newcommand{\mnras}{{Mon.~Not.~R.~Astron.~Soc.}}
\usepackage{graphicx}
\usepackage[large]{subfigure}
\usepackage{amssymb, amsmath}
\usepackage[amssymb]{SIunits}
\usepackage{epstopdf}
\usepackage{aas_macros}
\usepackage{natbib}

\newcommand{\ii}{a}
\newcommand{\jj}{b}

\newcommand{\kprof}{\underline{\kappa}}
\newcommand{\tprof}{\underline{\theta}}

\newcommand{\myvector}[1]{\bm{#1}}
\newcommand{\myunitvector}[1]{\bm{\hat{#1}}}

\begin{document}

\title{Accurate Analytic Model for the Weak Lensing Convergence One-Point Probability Distribution Function
       and its Auto-Covariance}
\author{Leander~Thiele\footnote{lthiele@princeton.edu}}
\affiliation{Perimeter Institute for Theoretical Physics, Waterloo ON N2L 2Y5, Canada}
\affiliation{Department of Physics, Princeton University, Princeton, NJ, USA 08544}
\author{J.~Colin Hill}
\affiliation{Department of Physics, Columbia University, 538 West 120th Street, New York, NY, USA 10027}
\affiliation{Center for Computational Astrophysics, Flatiron Institute, New York, NY, USA 10003}
\author{Kendrick~M.~Smith}
\affiliation{Perimeter Institute for Theoretical Physics, Waterloo ON N2L 2Y5, Canada}

\begin{abstract}
The one-point probability distribution function (PDF) is a powerful summary statistic
for non-Gaussian cosmological fields, such as the weak lensing (WL) convergence reconstructed from
galaxy shapes or cosmic microwave background (CMB) maps.
Thus far, no analytic model has been developed that successfully describes the
high-convergence tail of the WL convergence PDF for small smoothing scales from first principles.
Here, we present a halo-model formalism to compute the WL convergence PDF, building upon our previous results for the thermal Sunyaev-Zel'dovich field.
Furthermore, we extend our formalism to analytically compute the covariance matrix of the convergence PDF.
Comparisons to numerical simulations generally confirm the validity of our formalism in the non-Gaussian, positive tail of the WL convergence PDF, but also reveal the convergence PDF's strong sensitivity to small-scale
systematic effects in the simulations (e.g., due to finite resolution).
Finally, we present a simple Fisher forecast for a Rubin-Observatory-like survey,
based on our new analytic model.
Considering the $\{A_s, \Omega_m, \Sigma m_\nu\}$ parameter space and assuming a {\it Planck} CMB prior on
$A_s$ only, we forecast a marginalized constraint $\sigma(\Sigma m_\nu) \approx 0.08$ eV from the WL convergence PDF alone,
even after marginalizing over parameters describing the halo concentration-mass relation.
This error bar on the neutrino mass sum is comparable to the minimum value allowed in the normal
hierarchy, illustrating the strong constraining power of the WL convergence PDF.
We make our code publicly available at \url{https://github.com/leanderthiele/hmpdf}.
\end{abstract}

\maketitle


\section{Introduction}\label{sec:intro}

While the cosmic microwave background (CMB) historically has been the driving force in cosmological
parameter inference, we are now experiencing a proliferation in high-quality data from the late-time
matter distribution.
In contrast to the primary CMB, late-time fields are described by non-linear clustering of matter, rendering
the distribution of many relevant observables highly non-Gaussian.
For such non-Gaussian fields, the problem of extracting all information contained therein is unsolved;
while for Gaussian fields, such as the primary CMB, the power spectrum is an optimal summary
statistic containing all the information, no such summary statistic is known in the non-Gaussian
case.

One late-time field of interest is the weak lensing (WL) convergence. Weak gravitational lensing
describes the deflection of light by the matter distribution, imparting a shear and magnification on
the images of observed background galaxies or CMB fluctuations. The WL convergence is a redshift-weighted measure of the
integrated matter density along the line of sight; thus, it is a powerful probe of the matter
distribution.

Since in the course of the non-linear gravitational clustering the matter distribution departs
significantly from Gaussianity, appreciable amounts of information leak from the power spectrum into
higher-order statistics.
This motivated previous studies to consider parameter inference from such measures of
non-Gaussianity, such as the WL skewness and  bispectrum
\cite{Bernardeau1997WLstatistics,
      Schneider1998aperturemass,
      Waerbeke2001WLskewness,
      TakadaJain2004WLpowerspectrumbispectrum,
      KilbingerSchneider2005WLhigherorder,
      CoorayHu2000WLbispectrum,
      Huterer2006WLselfcalibration,
      Semboloni2013WLselfcalibration}.
An alternative summary statistic is the one-point probability distribution function (PDF), which
simply constitutes the histogram of WL convergence pixel values.
Originally considered in the context of peak statistics
\cite{Reblinsky1999peakstatistics,
      JainWaerbeke2000peakstatistics},
more recent studies have demonstrated that the WL convergence PDF can add significant constraining
power in parameter inference, not only in the $\sigma_8$-$\Omega_m$ plane,
but also on the neutrino mass sum
(e.g., Refs. \citep{Liu2018WLPDFneutrinomass,
                    Liu2016CMBlensingPDFConstraints,
		    Patton2016WLPDFconstraints}).

In some respects WL shares similarities with the thermal Sunyaev-Zel'dovich (tSZ) effect,
which describes the scattering of CMB photons by hot electrons residing mostly in massive halos.
Because the tSZ signal is approximately proportional to $M_\text{halo}^{5/3}$, the halo model allows
an excellent description of the tSZ PDF.
This fact was recently utilized in order to construct a semi-analytic model for the tSZ PDF
\cite{Hill2014, THS2019} (here ``semi-'' accounts for the fact that the model contains some functions
that are most accurately fixed by fitting to numerical simulations).

In this work, we demonstrate that the halo-model formalism developed in the tSZ context can be
applied to the WL convergence PDF as well, with small modifications due to two complications:
(1) In constrast to tSZ, the WL convergence signal does not include the additional $M_\text{halo}^{2/3}$ temperature
bias, which brings the distribution closer to Gaussian and renders the halo model slightly less
accurate;
(2) Furthermore, while the tSZ signal is strictly positive, the WL convergence receives negative contributions
from underdense regions (voids).
In contrast to previous works on the subject
\cite{CoorayHu2000WLbispectrum,
      Taruya2002WLPDFlognormal,
      Wang2002WLPDF,
      DasOstriker2005WLPDF},
our formalism is better suited to describe relatively large positive values of the convergence PDF
(which are sourced by massive halos), while performing less well in the only mildly non-linear regime
and especially at negative convergences.
As we will show, our formalism is also not very accurate when the convergence field is smoothed on
large angular scales.
Perturbative methods
\cite{Stebbins1996,
      Villumsen1996,
      Bernardeau1997WLstatistics,
      JainSeljak1997WLstatistics,
      Kaiser1998WL,
      Schneider1998aperturemass,
      Waerbeke1999WLsurvey},
and the large deviation statistics formalism developed in Refs.~\cite{Uhlemannetal2018,
                                                                      Barthelemyetal2020}
are better suited in such a situation.
In terms of physical input, our formalism is quite similar to the stochastic numerical method
developed in Refs.~\cite{KainulainenMarra2009WLPDFstochastic,
                         KainulainenMarra2011WLhalomodel,
			 KainulainenMarra2011WLturboGL}.

Besides a theoretical model for the expected form of the WL PDF, in order to do parameter inference
we also require a prescription for its statistical distribution.
While this distribution is non-Gaussian and difficult to compute, a first step is the computation of
the covariance matrix.
In terms of practical applications, the covariance matrix can be useful if the PDF is sufficiently
downsampled such that the likelihood can be computed in the Gaussian approximation, as was done in
Ref.~\cite{Hill2014}; alternatively, non-Gaussian inference methods such as likelihood-free
inference \cite{Alsing2018} can benefit from the covariance matrix as a starting point. 
In view of these potential applications, we generalize the halo-model formalism to compute not only
the one- but also the two-point PDF, the latter being sufficient for computation of the covariance
matrix.

The remainder of this paper is structured as follows.
In Sec.~\ref{sec:theory}, we present the theoretical part of this work, starting from the general
theory of weak gravitational lensing and proceeding to our halo-model formalism for the one- and
two-point PDFs. There, we also discuss the modifications to the formalism in comparison to the
tSZ case.
In Sec.~\ref{sec:resultsop}, we present various results obtained with our formalism for the
one-point PDF: A number of
calculations intended to build up intuition on the WL convergence PDF, and comparisons to two
independent sets of numerical simulations.
In Sec.~\ref{sec:resultstp}, we turn to the two-point PDF and the covariance matrix of the one-point
PDF. We perform a null test and compare the analytic covariance matrix to a large $N$-body simulation.
In Sec.~\ref{sec:fisher}, we utilize the previous results to produce a simple Fisher parameter
forecast.
We conclude in Sec.~\ref{sec:conclusions}.
Further analytic calculations useful in building up intuition are presented in
Appendix~\ref{app:analytic},
some details on the numerical evaluation of the formalism are collected in
Appendix~\ref{app:numerics}, and in Appendix~\ref{app:assumptions} we discuss the validity of
several approximations.

\section{Theory}
\label{sec:theory}

\subsection{Background}
\label{subsec:background}

Gravitational lensing distorts and magnifies the shapes of distant sources (e.g., galaxies or CMB
fluctuations) as a result of the projected gravitational potential of matter along the
line-of-sight (LOS), including dark and baryonic matter.  In the weak lensing limit, these effects are
encoded in the lensing convergence field, $\kappa(\myunitvector{n})$:
\begin{equation}
\label{eq.kappadef}
\kappa(\myunitvector{n}) = \int_0^{\infty} dz \, \delta(\myvector{x}(\chi(z)\myunitvector{n},z)) W^{\kappa}(z) \,,
\end{equation}
where $\chi(z)$ is the comoving distance to redshift $z$, $\delta$ is the matter density
fluctuation, $\delta(\myvector{x}) \equiv (\rho(\myvector{x}) - \bar{\rho})/\bar{\rho}$, and the lensing projection kernel is given by
\begin{equation}
\label{eq.Wkappa}
W^{\kappa}(z) = \frac{3}{2} \Omega_m H_0^2 \frac{(1+z)}{H(z)} \frac{\chi(z)}{c} \int_{z}^{\infty} dz_s \frac{dn}{dz_s} \frac{(\chi(z_s)-\chi(z))}{\chi(z_s)} \,,
\end{equation}
where $dn/dz_s$ is the distribution of sources, normalized such that $\int dz \, dn/dz = 1$.  Note
that for CMB lensing, $dn/dz = \delta_D(z-z_*)$, where $\delta_D$ is the Dirac-$\delta$ function and
$z_* \approx 1100$ is the redshift of last scattering.  For weak lensing due to galaxies, $dn/dz$ is
generally a more complicated function.  Note that we have specialized to the case of a flat universe
in Eqs.~(\ref{eq.kappadef}) and~(\ref{eq.Wkappa}).  For reference, the lensing convergence is
related to the lensing potential, $\phi(\myunitvector{n})$, via $\kappa(\myunitvector{n}) =
-\nabla^2 \phi(\myunitvector{n}) / 2$ (where $\nabla^2$ is the two-dimensional Laplacian on the sky), or $\kappa_{\ell}
= \ell(\ell+1) \phi_{\ell}/2$ in harmonic space.

Given a 3D halo density profile (e.g., the NFW profile~\cite{NFW1997}),
we can define the lensing convergence profile, $\kprof(\myvector{\theta}, M, z)$
for a halo of mass $M$ at redshift $z$:
\begin{equation}
\label{eq.kappaprofile}
\kprof(\myvector{\theta}, M, z) = \Sigma_{\rm crit}^{-1}(z) \int_{\rm LOS} \rho \left( \sqrt{l^2 +
d_A^2 |\myvector{\theta}|^2}, M, z \right) dl \,,
\end{equation}
where $\rho(\myvector{r},M,z)$ is the halo density profile, $d_A(z)$ is the angular diameter distance to redshift $z$, and $\Sigma_{\rm crit}(z)$ is the critical surface density (in physical units here) for lensing at redshift $z$ assuming a source distribution $dn/dz$:
\begin{eqnarray}
\label{eq.Sigmacrit}
\Sigma_{\rm crit}^{-1}(z) & = & \frac{4 \pi G \chi(z)}{c^2 (1+z)} \int_z^{\infty} dz' \frac{ \left(\chi(z')-\chi(z) \right)}{\chi(z')} \frac{dn}{dz'} \\
 & = & \frac{8 \pi G}{3 \Omega_m H_0^2} \frac{H(z)}{c (1+z)^2} W^{\kappa}(z) \,.
\end{eqnarray}
For a spherically symmetric density profile, the convergence profile is azimuthally symmetric,
i.e., $\kprof(\myvector{\theta}, M, z) = \kprof(\theta, M, z)$.
For the NFW density profile, analytic forms exist for the convergence profile.  One subtlety, however, is the non-convergence of the enclosed mass in the NFW profile as $r \rightarrow \infty$, which thus necessitates a radial cutoff in calculations using this profile.

\subsection{WL PDF in the Halo Model}
\label{subsec:PDF}

In Ref.~\cite{THS2019}, an analytic approach based on the halo model was constructed to describe the one-point PDF of the tSZ field,
building on a simpler model presented in Ref.~\cite{Hill2014}.
In particular, the effects of halo overlaps along the LOS and halo clustering, which were neglected
in Ref.~\cite{Hill2014}, were included in Ref.~\cite{THS2019}.
However, the expressions derived in Ref.~\cite{THS2019} are more broadly applicable to the one-point PDF
of {\it any} (projected) cosmic field that can be modeled in a halo-based approach.
The halo model approach is very accurate for the tSZ field
as this field is heavily dominated by contributions from massive halos~(e.g., Refs.~\cite{KomatsuSeljak2002tSZpowerspectrum,Battaglia2012,HillPajer2013tSZpowerspectrum}),
due to the temperature dependence of the tSZ signal.
A primary goal of this paper is to assess the accuracy of this model for other cosmic fields, in particular the WL convergence field.
Thus, as a first step to model the WL convergence PDF, we can simply use the expressions from~\cite{THS2019},
but with the $y$ (tSZ) profile replaced by the $\kappa$ profile defined in Eq.~\eqref{eq.kappaprofile}.
The rest of the formalism derived in that work then goes through unchanged.

For completeness and ease of reference in later sections, we include the derivation of the one-point $\kappa$ PDF in this formalism here.
In some places, algebraic manipulations are omitted for brevity; we refer the interested reader to
Ref.~\cite{THS2019} for full details. 

\subsubsection{One-Halo Term}
\label{subsubsec:onehalo}

We refer to the (differential) $\kappa$ one-point PDF as $P(\kappa)$.  Considering a bin spanning $[\kappa_i, \kappa_{i+1}]$, we define the binned version of the PDF as
\begin{equation}
p_i = \int_{\kappa_i}^{\kappa_{i+1}} d\kappa \, P(\kappa) \,.
\end{equation}
The fundamental concept underlying the model developed in Refs.~\cite{Hill2014,THS2019} is that $p_i$
quantifies the sky fraction subtended by $\kappa$ values in the range $[\kappa_i,\kappa_{i+1}]$.
For an individual spherically symmetric halo with an azimuthally symmetric projected
$\kappa$-profile $\kprof(\theta)$, this sky fraction is simply the area in the annulus between $\theta(\kappa_i)$ and $\theta(\kappa_{i+1})$, where $\theta(\kappa_i)$ is the angular distance from the center of the profile to the radius where $\kappa(\theta) = \kappa_i$.  If we then assume that halos are sufficiently rare that they never overlap on the sky, the one-point PDF is simply given by adding up the annular area contributions from all halos:
\begin{equation}
\label{eq:H14}
p_i = \int dz \, dM \frac{\chi^2}{H} \frac{dn}{dM} \pi \left( \tprof^2(\kappa_{i}) - \tprof^2(\kappa_{i+1}) \right) + \delta_i\,(1-F_{\rm halos}) \,,
\end{equation}
where $dn(M,z)/dM$ is the halo mass function (i.e., the number of halos of mass $M$ at redshift $z$
per unit mass and comoving volume), $\tprof(\kappa,M,z)$ is the inverse function of $\kprof(\theta,M,z)$, $F_{\rm halos}$ is the total sky area subtended by all halos (assuming some radial cutoff for the halo profiles), $\delta_i$ is unity if $\kappa=0$ lies in the bin and zero otherwise, and redshift and mass dependences have been suppressed in the equation for compactness.
Equation~\eqref{eq:H14} is only accurate in the limit in which halos do not overlap on the sky; moreover, it neglects effects due to the clustering of halos.  While these assumptions are (moderately) accurate for the tSZ field, they are not accurate for the WL convergence field.

We thus seek a more general approach, in which these limiting assumptions are discarded.  The basic
ideas and results for our improved formalism were presented for the tSZ field in Ref.~\cite{THS2019}; here, we adapt the formalism to the WL convergence field and introduce a more compact notation.
Our goal is to compute the one- and two-point PDFs, $P(\kappa_\ii)$ and $P(\kappa_\ii, \kappa_\jj; \phi)$,
where $\phi$ is the angular separation between two sky locations at which we measure the two convergence
values $\kappa_\ii$ and $\kappa_\jj$. It will be convenient to work in Fourier space, introducing
\begin{eqnarray}
P_{\ii}    &\equiv& \int d\kappa_\ii e^{i\lambda_\ii\kappa_\ii} P(\kappa_\ii)\,, \\
P_{\ii\jj} &\equiv& \int d\kappa_\ii d\kappa_\jj e^{i(\lambda_\ii\kappa_\ii +
\lambda_\jj\kappa_\jj)} P(\kappa_\ii, \kappa_\jj; \phi)\,,
\end{eqnarray}
where we have abbreviated the notation for conciseness.

We will separate the PDFs into a one- and a two-halo term, writing
$P = P^\text{1h} P^\text{2h}$. In this section we compute the 1-halo term, i.e., we ignore the
clustering of halos for the moment.
We introduce two further pieces of notation: we denote the projected halo mass function
\begin{equation}
n\equiv n(M,z) = \frac{\chi^2(z)}{H(z)}\frac{dn(M,z)}{dM}\,,
\end{equation}
which gives the expected number of halos per unit mass and redshift interval in unit solid angle;
furthermore we introduce the quantities
\begin{eqnarray}
\hat{K}^{(\theta)}_\ii &\equiv& e^{i\kprof(\theta)\lambda_\ii} - 1\,, \\
K^{(\ell)}_\ii         &\equiv& \int_\theta\hat{K}^{\myvector{(\theta)}}_\ii
J_0(\ell\theta)\,,\label{eq:Kiell}
\end{eqnarray}
where $\kprof$ are the convergence profiles, $\int_\theta\equiv\int 2\pi\theta d\theta$
and we have again suppressed the mass and redshift dependences.

First, we consider a narrow bin of width $dM\,dz$ in mass-redshift space, such that halo overlaps can be
neglected for the infinitesimal number of halos in this bin.
For a given realization of the halo distribution, we have at the arbitrarily chosen origin
$\myvector{0}$
\begin{equation}
e^{i\lambda_\ii\kappa(\myvector{0})} = 1 + \sum_h \hat{K}^{(\theta_h)}_\ii\,,
\end{equation}
where the sum runs over all halos in the given mass-redshift bin and $\theta_h$ is their separation
from the origin.
Thus, we find for the 1-point PDF
\begin{eqnarray}
P^\text{1h}_\ii &=& \left\langle e^{i\lambda_\ii\kappa(\myvector{0})} \right\rangle_h \nn\\
		&=& 1 + dM\,dz\,n \int_{\myunitvector{n}} \hat{K}^{(\myunitvector{n})}_\ii \nn\\
                &=& 1 + dM\,dz\,n K^{(0)}_\ii\,,
\end{eqnarray}
where the subscript $h$ indicates that we are averaging over realizations of the halo distribution.
Likewise, we find for the 2-point PDF
\begin{eqnarray}
P^\text{1h}_{\ii\jj} &=& \left\langle
                         e^{i\lambda_\ii\kappa(\myvector{0})}e^{i\lambda_\jj\kappa(\myvector{\phi})}
		         \right\rangle_h \nn\\
		     &=& 1 + dM\,dz\,n \left[K^{(0)}_\ii+K^{(0)}_\jj
		                             +\int_{\myunitvector{n}}\hat{K}^{(\myunitvector{n})}_\ii\hat{K}^{(\myunitvector{n}-\myvector{\phi})}_\jj\right]\nn\\
		     &=& P^\text{1h}_\ii P^\text{1h}_\jj\left[1+dM\,dz\,n\int_\ell
		         K^{(\ell)}_\ii K^{(\ell)}_\jj J_0(\ell\phi)\right]\,,
\end{eqnarray}
where we have introduced $\int_\ell\equiv\int\ell d\ell/2\pi$.
Note that to the order considered so far, terms of the form $1+dM\,dz\,A$ can equally well be written
as $\exp dM\,dz\,A$.
Under the Born approximation, the convergence is an additive quantity. Thus, the complete PDFs can
be obtained by convolution, which is equivalent to multiplication in Fourier space:
\begin{eqnarray}
P^\text{1h}_\ii &=& \exp\int_{M,z} K^{(0)}_\ii\,,\label{eq:onePDFonehalo}\\
\frac{P^\text{1h}_{\ii\jj}}{P^\text{1h}_\ii P^\text{1h}_\jj} &=&
                \exp\int_{M,z,\ell}K^{(\ell)}_\ii K^{(\ell)}_\jj J_0(\ell\phi)\,,\label{eq:twoPDFonehalo}
\end{eqnarray}
where we have introduced $\int_M\equiv\int dM\,n$, $\int_z\equiv\int dz$ for brevity.
As we have demonstrated in Ref.~\cite{THS2019}, expanding the exponentials to first order leads to
the approximate model from Ref.~\cite{Hill2014}. In this sense, terms of order $n^p$ in the Taylor 
expansion of $P^\text{1h}$ can be interpreted as describing overlaps of $p$ halos along the line of
sight.

\subsubsection{Two-Halo Term}
\label{subsubsec:twohalo}

The two-halo term arises from the dependence of halo density on the underlying long-wavelength
linear density field, which at sky location $\myunitvector{n}$ and redshift $z$ we denote by
\begin{equation}
\delta(\myunitvector{n}) \equiv \delta_\text{lin}(\myunitvector{n},z)\,.
\end{equation}
The change in the halo density can to a first approximation be written as
\begin{equation}
n \rightarrow n(\myunitvector{n}) = n[1 + b\delta(\myunitvector{n})]\,,
\label{eq:biasdef}
\end{equation}
where $b\equiv b(M,z)$ is the linear halo bias.
In order to compute the 2-halo term in the PDFs, we proceed in two steps: first, we compute the
correction to the 1-halo term in a given realization $\delta$, obtaining $P^\delta$, and then we
perform the average over realizations.
We note that in contrast to Ref.~\cite{THS2019} we denote by $P^\delta$ only the multiplicative
correction factor to the PDF.
Using the substitution in Eq.~\eqref{eq:biasdef}, the required correction factors can be written down
immediately:
\begin{eqnarray}
P^\delta_\ii(\myvector{0}) &=& \exp\int_z\alpha_\ii\delta(\myvector{0})\,,\\
P^\delta_{\ii\jj}(\myvector{0}) &=&
           \exp\int_z\alpha_\ii\delta\big(-\frac{\myvector{\phi}}{2}\big)
	            +\alpha_\jj\delta\big(\frac{\myvector{\phi}}{2}\big)
		    +\beta_{\ii\jj}\delta(\myvector{0})\,,
\end{eqnarray}
where we have introduced
\begin{eqnarray}
\alpha_\ii     &\equiv& \int_M b K^{(0)}_\ii\,, \\
\beta_{\ii\jj} &\equiv& \int_{M,\ell} b K^{(\ell)}_\ii K^{(\ell)}_\jj J_0(\ell\phi)\,.
\end{eqnarray}
We take the opportunity to point out a subtlety here: because we are working with a fixed
realization $\delta$ at the moment, isotropy is broken and the PDFs depend explicitly on sky
location. Thus, we need to assume that the linear density field $\delta$ varies sufficiently slowly
that the halo model formalism we have been assuming still makes sense.
As we will explicitly demonstrate in Appendix~\ref{subapp:powerspectrum}, this assumption is equivalent
to the statement that the linear matter correlation function vanishes on scales similar to typical
halo radii.
Now all that is left to do is to perform the average over realizations of $\delta$,
giving the clustering corrections
\begin{equation}
P^\text{2h} = \langle P^\delta \rangle_\delta\,.
\end{equation}
We remind the reader that for light-cone integrals $f_i(\myunitvector{n})=\int_z W_i
\delta(\myunitvector{n})$
the Limber approximation allows us to write
\begin{equation}
\left\langle f_i(\myvector{0})f_j(\myvector{\phi})\right\rangle_\delta
     = \int_z H \zeta(\phi) W_i W_j\,,
\end{equation}
where $\zeta(\phi)$ is the redshift-dependent line-of-sight projected matter correlation function,
which in terms of the linear matter power spectrum can be written as
\begin{equation}
\zeta(\phi) = \int\frac{k\,dk}{2\pi} P_\text{lin}(k,z) J_0(k\phi\chi(z))\,.
\end{equation}
Utilizing the Limber approximation and the identity $\langle e^x\rangle = e^{\langle x^2\rangle/2}$
valid for Gaussian distributed $x$, we obtain
\begin{eqnarray}
P^\text{2h}_\ii  &=& \exp\int_z\frac{H\zeta(0)}{2}\alpha_\ii^2\,,\label{eq:onePDFtwohalo}\\
\frac{P^\text{2h}_{\ii\jj}}{P^\text{2h}_\ii P^\text{2h}_\jj} &=& 
       \exp\int_z H \big[\alpha_\ii\alpha_\jj\zeta(\phi)
                          +\frac{1}{2}\beta_{\ii\jj}^2\zeta(0) \nn\\
&&\hspace{1.5cm}
+\beta_{\ii\jj}(\alpha_\ii+\alpha_\jj)\zeta\big(\frac{\phi}{2}\big)\big]\,.\label{eq:twoPDFtwohalo}
\end{eqnarray}
This concludes the main theoretical part of this work. We would like to point out two features of
the formalism presented here:
(1) Although we have assumed that halos are only described by their mass and redshift, one could
consider additional labels $\myvector{c}$ (e.g., related to halo environment or formation history).
This would introduce a $\myvector{c}$-dependence in the halo
mass function $n$ and add further integrations over $\myvector{c}$ on equal footing with the mass
integrations (the redshift integrations are special because of the simplification introduced by the
Limber approximation);
(2) The general formalism applies to any $N$-point PDF. For example, for the 3-point PDF one would
have to compute three-index objects $\gamma_{ijk}$ that are analogous to $\alpha_i$ and
$\beta_{ij}$. However, the ``momentum" labels $\ell$ would turn into vectorial quantities now,
which presumably complicates the required integrations considerably.

\subsection{WL PDF Contributions from Non-Virialized Matter and Voids}
\label{subsec:nonvirial}

The expressions above only account for the contributions to the WL $\kappa$ PDF
due to matter in halos (``virialized matter'').
For the Compton-$y$ field, this approximation was very accurate
due to the temperature dependence of the tSZ signal,
which strongly biases the $y$ field toward electrons in massive halos.
For $\kappa$, this approximation is less accurate,
as the WL convergence field is an unbiased tracer of the matter distribution.
Moreover, in contrast to Compton-$y$,
there are negative-signal regions in the $\kappa$ field
(i.e., projected under-densities in the matter distribution).
We thus require some method to treat both the ``matter outside of halos'' and ``voids.''

With regard to the matter outside of halos, one option would be the following.
We assume that the rest of the $\kappa$ map (not accounted for by the halo-based model)
is purely a Gaussian random field (GRF) that is uncorrelated with the virialized-halo part of the $\kappa$ map.
We can compute the variance of this GRF (call it the ``residual variance'')
by simply using the halo model to compute the angular power spectrum $C_{\ell}^{\kappa\kappa}$ and truncating the halo model
integrals at the $M$ values above which the explicit profile-based calculation is used
(so that the variance contributed by those objects is not double-counted).  Alternatively, we could
compare the variance of the halo-model PDF with the variance obtained from 
the ``Halofit'' fitting function~\cite{Smith2003Halofit,Takahashi2012Halofit}, and extract the residual variance from the difference of these quantities.
After computing the residual variance,
a Gaussian PDF of this width can be convolved with the halo-based PDF to obtain the final $\kappa$ PDF.

We implement both approaches described above.
We find the changes in the PDF with respect to the unmodified halo-model-only result to be extremely minor.
The first approach suffers from the problem that the halo model becomes ill-defined in the low-mass regime relevant to this calculation, and thus the small residual variance is quite uncertain.
On the other hand, we frequently find the variance computed from Halofit to be smaller than the variance
deduced from the halo-model PDF, invalidating the basic assumption.  Thus, in this work we do not include either of these ideas for incorporating convergence contributions from matter outside of halos, instead using only the halo model described in the previous section.  If exact results are necessary, we recommend testing stability with respect to the lower limit in mass integrations.

With regard to the negative-$\kappa$ voids, a simple prescription is based on the fact
that the mean $\langle\kappa\rangle = 0$ by construction.
Thus, to a first approximation, we simply compute the halo-model PDF as described above and then
shift it such that the physical constraint is enforced.
This idea clearly fails to provide an accurate description of the negative-$\kappa$ tail of the PDF.
It also is not immediately obvious that it leads to good predictions for the positive-$\kappa$
tail, primarily because it does not take into account void-halo correlations.
Thus, comparison to numerical simulations will be crucial in assessing the accuracy
of this simple approximation.

\section{Results: One-point PDF}
\label{sec:resultsop}

Before discussing various results obtained with the formalism developed in the previous section,
we mention several choices for fitting functions and numerical settings.
For the halo mass function $dn/dM$ and the linear bias $b$, we use the fitting functions of
Ref.~\cite{Tinker2010}.
We describe halos with an NFW profile, using the concentration-mass relation of
Ref.~\cite{Duffy2008}.
We use the Colossus package~\cite{Diemer2018Colossus} for calculations in the halo model,
and CAMB~\cite{CAMB} and CLASS~\cite{CLASS1,CLASS2}
(with Halofit corrections from Refs.~\cite{Smith2003Halofit,Takahashi2012Halofit})
for matter and WL convergence power spectra.
As mentioned before, the NFW profile necessitates a radial cutoff; we find the WL convergence PDF to
be nearly independent of this cutoff and choose it at $r_\text{max} = 1.6 r_\text{vir}$, where
$r_\text{vir}$ is computed according to Ref.~\cite{BryanNorman1998}
and the prefactor of 1.6 is chosen to obtain good agreement
between the halo-model-computed and Halofit-computed WL convergence power spectra.
Unless otherwise stated, all halo masses are given in terms of $M\equiv M_\text{200c}$,
and we choose integration limits $11\leq\log_{10} M/h^{-1}M_\odot\leq 16$, so that the PDF is very well
converged (as will be shown in Sec.~\ref{subsec:MZcontributions}).
For simplicity, we specialize to a Dirac-$\delta$ distribution of source galaxies, $dn/dz = \delta_D(z-z_s)$,
at a single source redshift $z_s$.
In order to incorporate pixelization effects, we convolve the convergence profiles with a window
function
\begin{equation}
\label{eq.pixelwindow1D}
W_{\text{1pt},\ell}^\text{pix} = \frac{4}{\pi}\int_0^{\pi/4} d\varphi\,
         \text{sinc}(\cos(\varphi)\ell a)\text{sinc}(\sin(\varphi)\ell a)\,,
\end{equation}
where $a$ is half the pixel side length.
This prescription is only approximate: there is no precise method to incorporate quadratic pixels while
keeping the convergence profiles azimuthally symmetric.
However, the error incurred is negligible for the purposes of this work.
Note that it would also be irrelevant in any real parameter inference,
since for realistic shape noise levels the Wiener filter one would apply to the map
(as described in Sec.~\ref{sec:fisher}) cuts off harmonic-space modes before the pixelization effect
becomes relevant.

Having developed the analytic formalism in the previous section, we now proceed to discuss various results in the following subsections.
In Sec.~\ref{subsec:overlapsclustering}, we examine the effect of corrections
(overlaps and clustering) on the WL one-point PDF.
In Sec.~\ref{subsec:comptosims}, we compare our model's predictions to results from two sets of
cosmological $N$-body simulations.
In Sec.~\ref{subsec:MZcontributions}, we disentangle the contributions from different halo mass and
redshift intervals to the PDF.
In Sec.~\ref{subsec:parameterdependence}, we discuss the dependence of the WL PDF on cosmological and concentration model parameters.

\subsection{Impact of Overlaps and Clustering}
\label{subsec:overlapsclustering}

\begin{figure}
	\includegraphics[width=0.5\textwidth]{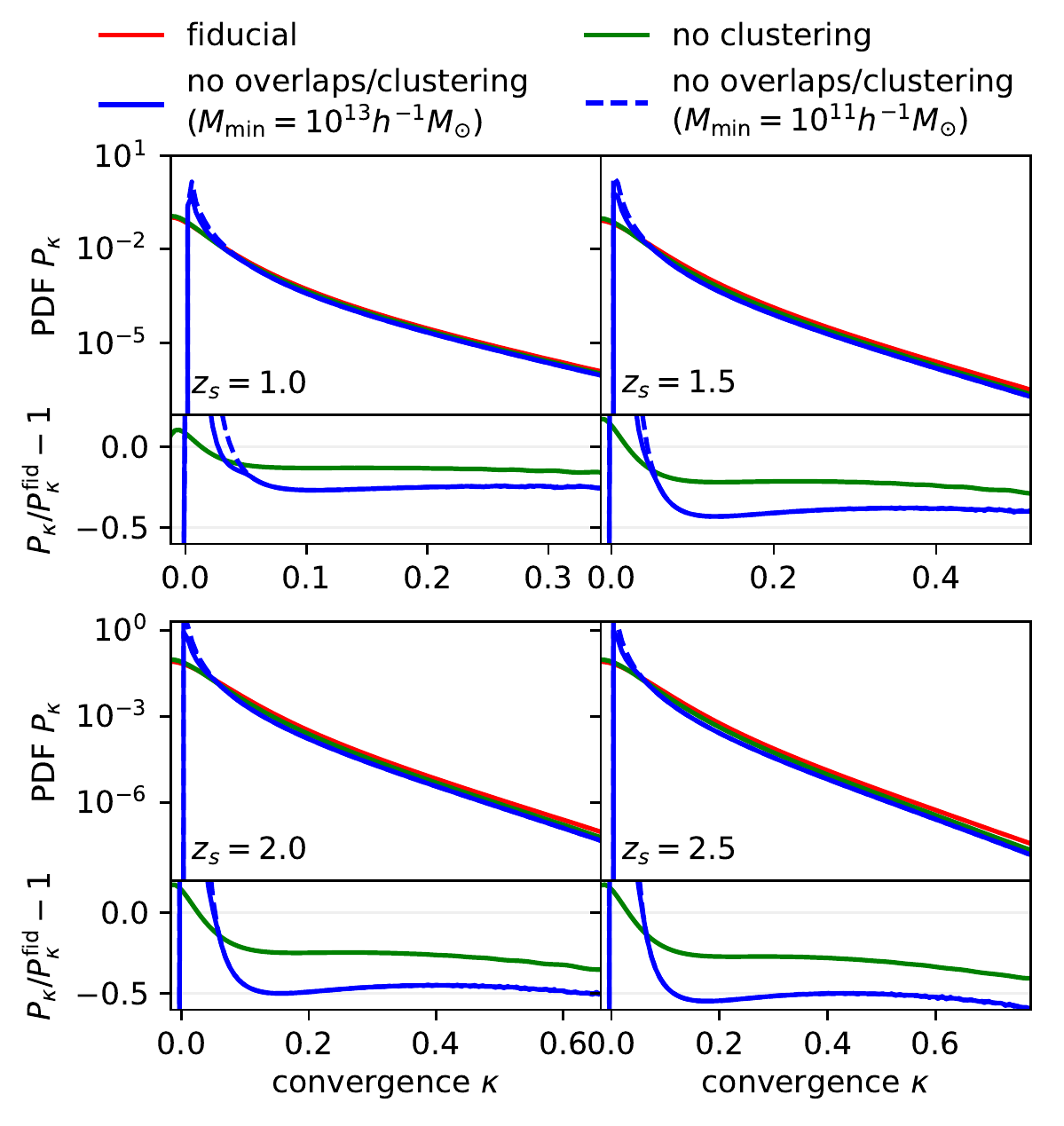}
	\caption{
	The effects of clustering and overlaps for four different source redshifts $z_{\text s}$.
	The PDF neglecting clustering (\emph{green}) seems to show similar behavior to what was
	observed in the tSZ case~\cite{THS2019}, but the clustering effect is much more pronounced.
	Note that the shift to $\langle\kappa\rangle=0$ does not make sense in the no-overlaps
	formalism: by assumption, the sky is infinitely large so the condition is automatically
	satisfied (this corresponds to the divergence at $\kappa=0$).
	}
	\label{fig:overlapsclustering}
\end{figure}

As discussed before, our formalism utilizes a halo-model-based framework similar to the tSZ PDF
calculation in Ref.~\cite{Hill2014}, with the crucial difference that we incorporate corrections
arising from halo clustering and overlaps along the line of sight (as developed in Ref.~\cite{THS2019}).
Naturally, we should examine the size of these corrections.
In Fig.~\ref{fig:overlapsclustering}, we plot the exact result from our formalism in red, while
the result neglecting halo clustering is represented in green.
As noted in Ref.~\cite{Hill2014}, the result neglecting overlaps is only applicable if the minimum
halo mass contribution is relatively large, thus, we plot two versions of the PDF neglecting both
overlaps and clustering with different minimum masses in solid/dashed blue.
These results are shown for four different choices of source redshift ranging from $z_s = 1$ to 2.5.

We see that both clustering and overlaps constitute substantial corrections, of order a few $10\,\%$.
This is in marked contrast to the conclusion we drew in the tSZ case, where the clustering effect
was subdominant and did not exceed a few percent.
As we shall see in Sec.~\ref{subsec:MZcontributions}, the convergence PDF receives larger
contributions from low-mass halos at lower redshifts in comparison to the tSZ PDF, which explains
the more pronounced clustering contribution.
We note the unphysical divergence of the results neglecting overlaps near $\kappa = 0$, which is
removed by the improved model presented in this work.

\subsection{Comparison to Numerical Simulations}
\label{subsec:comptosims}

We compare results from our formalism to WL convergence PDFs extracted from two different sets of $N$-body simulations, namely
MassiveNuS\footnote{\url{http://astronomy.nmsu.edu/aklypin/SUsimulations/MassiveNuS/}}~(Sec.~\ref{subsubsec:MassiveNuS})~\cite{Liu2018MassiveNuS}
and Takahashi {\it et
al.}\footnote{\url{http://cosmo.phys.hirosaki-u.ac.jp/takahasi/allsky_raytracing/}}~(hereafter
T17, Sec.~\ref{subsubsec:T17})~\cite{Takahashi2017sims}.
Both of these simulations provide ray-traced WL convergence maps.  We provide further details on each simulation analysis below.

\subsubsection{Comparison to MassiveNuS}
\label{subsubsec:MassiveNuS}

\begin{figure}
	\includegraphics[width=0.5\textwidth]{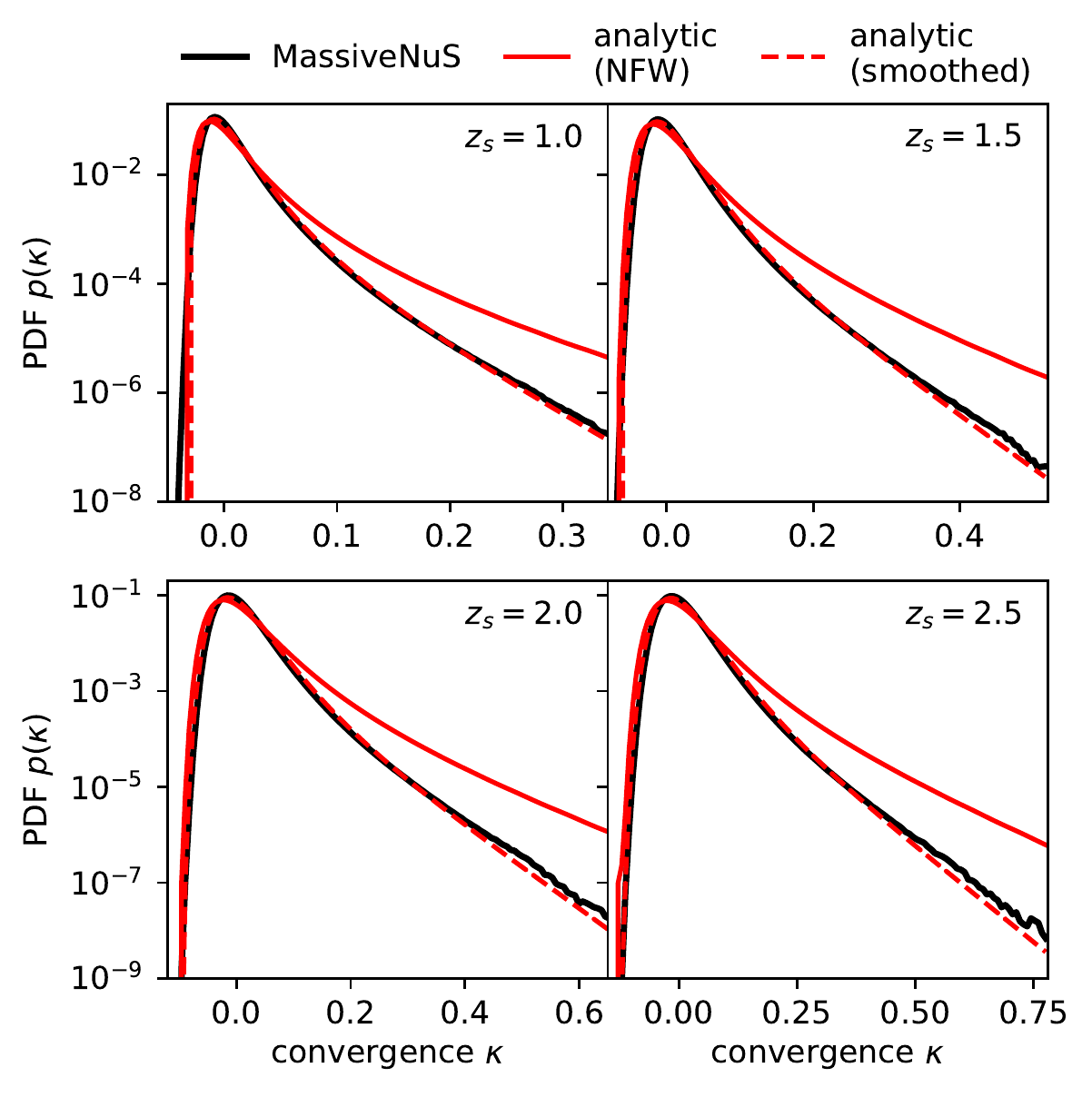}
	\caption{Comparison between the MassiveNuS simulations results
	         (\emph{black}) \cite{Liu2018MassiveNuS} and our analytic model (\emph{red})
		 for the WL convergence one-point PDF, for four different source redshifts.
	         \emph{Solid red}: fiducial result of our model;
	         \emph{dashed red}: the result obtained by smoothing the NFW
	         density profiles with a $k$-space filter calibrated on the convergence
	         power spectra measured in MassiveNuS, as described in the text
		 and illustrated in Fig.~\ref{fig:psMassiveNuS}.
		 This filter captures the non-negligible effects due to the finite resolution of the simulation.
		 Here, as well as in the other plots in which we show simulation data,
		 the error bars would be invisible by eye.}
	\label{fig:opMassiveNuS}
\end{figure}

\begin{figure}
	\includegraphics[width=0.5\textwidth]{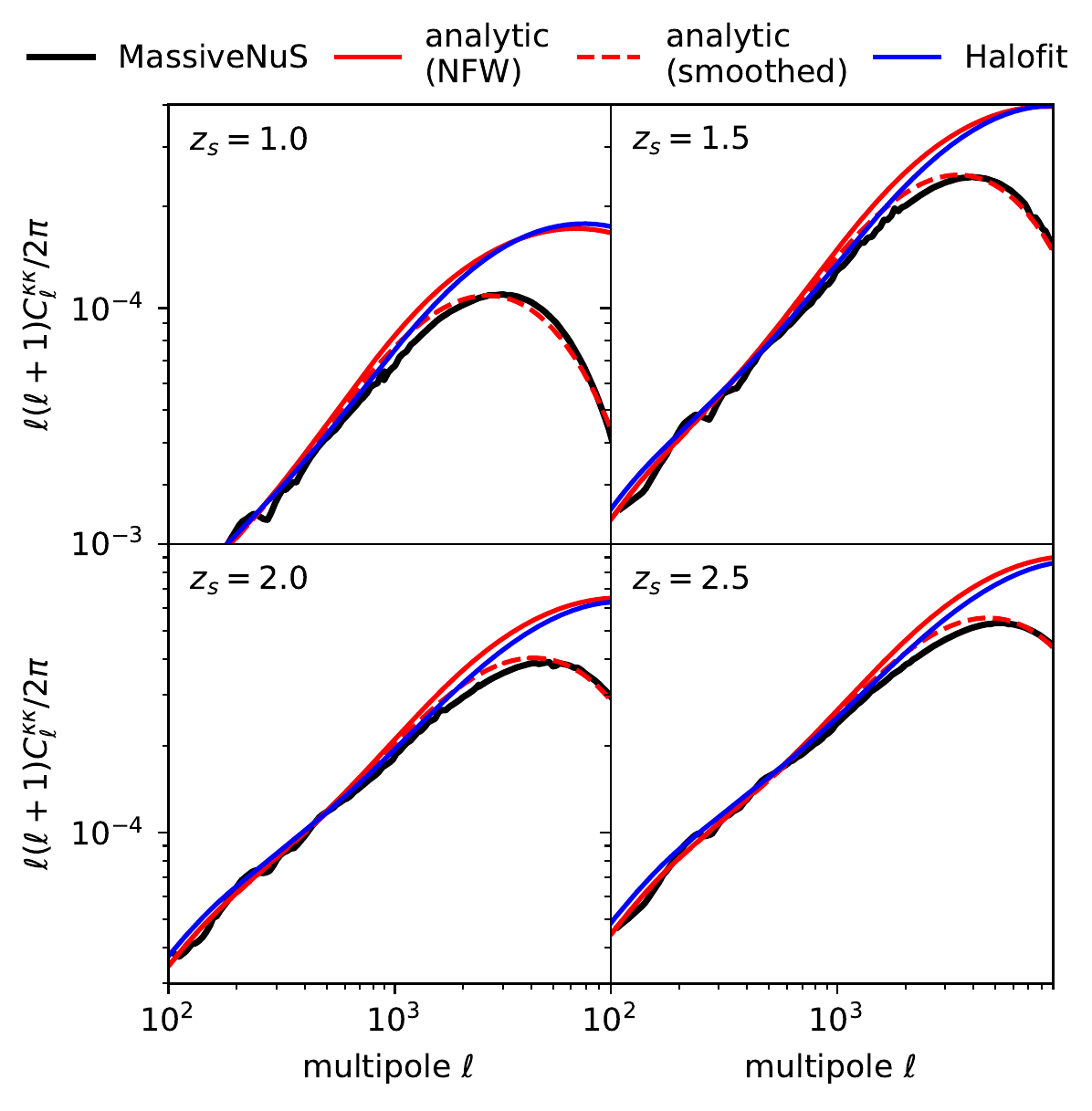}
	\caption{Convergence power spectra. \emph{Blue} is the Halofit result,
	         while \emph{solid red} is our fiducial analytic result from the halo model.
		 The MassiveNuS power spectra (\emph{black}) show a deficiency in power
		 at $\ell\gtrsim 10^3$, likely due to resolution effects.
		 The \emph{dashed red} lines are analytic power spectra obtained by
		 smoothing the NFW density profiles, as described in the text,
		 in order to mimic the resolution effect.}
	\label{fig:psMassiveNuS}
\end{figure}

We analyze a set of $10^4$ ray-traced weak lensing convergence maps from the MassiveNuS simulation
suite, which are derived from a set of $N$-body simulations that include dark matter and an
approximate treatment of massive neutrinos via a linear response method~\cite{Liu2018MassiveNuS}.
Each convergence map is $3.5 \times 3.5$ deg${}^2$ with $512^2$ square pixels, corresponding to a
pixel side length of $0.41$ arcmin.  The effect of the pixel window is treated in our analytic
calculations via Eq.~\eqref{eq.pixelwindow1D}.  The simulations include maps for a wide range of cosmological parameters, but we consider only the fiducial simulated cosmology, with parameters given by $\Omega_m = 0.3$, 
$\Omega_b = 0.046$,
$h = 0.7$,
$A_s = 2.1 \times 10^{-9}$,
$n_s = 0.97$, and zero neutrino mass ($\sigma_8 = 0.8523$ is a derived parameter).  We use these parameters in all analytic calculations that compare to MassiveNuS.

We analyze convergence maps constructed with $\delta$-function source planes at various redshifts.  We consider $\kappa$ values ranging from $[-5\sigma_{\kappa}, 20\sigma_{\kappa}]$, where $\sigma_{\kappa}$ is the variance of the $\kappa$ field measured from the full simulation set for each source redshift option.  The bins are linearly spaced with width $\sigma_{\kappa}/5$.  In all PDF measurements, we enforce the constraint that $\langle \kappa \rangle = 0$.

To ensure that the simulation results are robust to cosmic variance fluctuations resulting from the small map size, we also analyze a set of $10^5$ convergence maps that were produced for covariance matrix estimation at the fiducial cosmology, using additional, independent $N$-body simulations.  We sub-divide this large set into 10 subsets of $10^4$ maps each, and verify that any fluctuations in the measured $\kappa$ PDFs across the subsets are negligible.

In Fig.~\ref{fig:opMassiveNuS}, we plot a comparison between the fiducial analytic one-point PDF
(solid red) and the PDF measured in MassiveNuS (black), for four different source redshifts, $z_s = 1, 1.5, 2, 2.5$.
While the discrepancies at negative $\kappa$ are entirely expected, the large differences in the
positive-$\kappa$ tail are not expected given our intuition that the halo model should perform
very well in this regime.
In order to explain these discrepancies, we plot WL convergence power spectra in Fig.~\ref{fig:psMassiveNuS}.
We observe good agreement between the Halofit result (blue) and the fiducial result (solid red)
computed using the standard halo model expressions (e.g.,
Ref.~\cite{CoorayHuMiralda2000WLpowerspectrumhalomodel}).
On the other hand, MassiveNuS lacks power for $\ell\gtrsim 10^3$.
This is likely related to small-scale resolution effects in the simulation, presumably a combination
of finite mass resolution and force softening
(e.g., Refs.~\citep{ZorillaMatilla2019, JoyceGarrisonEisenstein2020}).
As a simple test of whether these resolution effects can explain the discrepancies seen in the
one-point PDFs, we calibrate a $k$-space filter with which we smooth the NFW density profiles
such that the resulting convergence power spectra match the MassiveNuS results.
We find the filter
\begin{equation}
W(k) = [ 1 + (kR)^2 ]^{-0.7}\,,
\end{equation}
where $R = 0.17\,h^{-1}\text{Mpc}$ comoving, to yield relatively good agreement.
The resulting power spectra are plotted in dashed red in Fig.~\ref{fig:psMassiveNuS}.
Having calibrated the filter $W(k)$ on the power spectra, we then compute the resulting one-point
PDF, plotted in dashed red in Fig.~\ref{fig:opMassiveNuS}.
We observe much better agreement now. In the part of the PDF that is relatively close to Gaussian
the analytic result matches the simulations almost exactly.
Small discrepancies remain in the high-$\kappa$ tail, which is not surprising since our smoothing filter
was calibrated solely on the two-point correlation function, while the one-point PDF in the tail
depends strongly on higher-order correlation functions.
Thus, we conclude that the resolution effects leading to a lack of power at high $\ell$ are likely
responsible for the high-$\kappa$ discrepancy between our model and the simulation result, rather than a deficiency of our halo model formalism.

\subsubsection{Comparison to T17}
\label{subsubsec:T17}

\begin{figure}
	\includegraphics[width=0.5\textwidth]{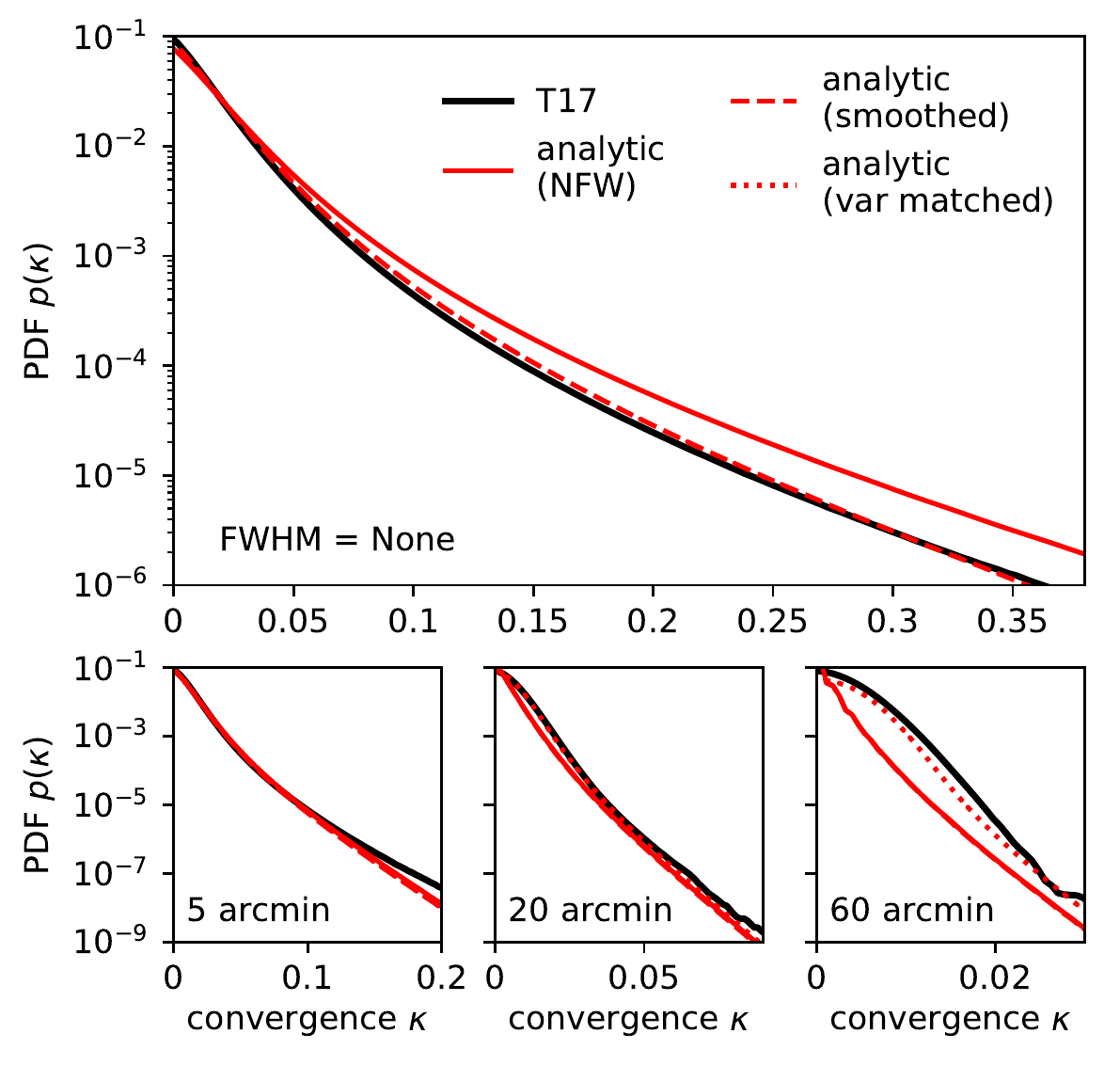}
	\caption{Comparison between the T17 simulation results
	         \citep[black]{Takahashi2017sims} and the analytic result (red)
		 for the convergence one-point PDF, for four different Gaussian smoothing scales labeled by their FWHM in the panels.
		 \emph{Solid red}: fiducial result of our model;
		 \emph{dashed red}: the result obtained by smoothing the NFW density profiles;
		 \emph{dotted red}: the result obtained by convolving with a Gaussian such that the
		 final variance matches the variance measured in the simulation.
		 Note that in the top and bottom left panels the analytic variance is slightly
		 larger than the one measured in the simulations,
		 thus no dotted line is plotted.}
	\label{fig:opT17}
\end{figure}

We analyze a set of 108 full-sky, ray-traced weak lensing convergence maps from T17~\cite{Takahashi2017sims}, which are derived from a suite of large dark-matter-only $N$-body simulations.  The maps are provided in HEALPix~\cite{HEALPix2005} format at resolution $N_{\rm side} = 8192$, corresponding to an approximate pixel scale of $0.43$ arcmin.  The parameters used in the simulations are $\Omega_m = 0.279$, 
$\Omega_b = 0.046$,
$h = 0.7$,
$\sigma_8 = 0.82$,
$n_s = 0.97$, and zero neutrino mass.  We match these parameters in all analytic calculations that compare to T17.

In our comparison to the T17 simulations we focus on the effect of smoothing the convergence maps.
Thus, we only produce results for a single source redshift, $z_s = 1.0334$, but apply Gaussian filters
of varying full-width half-maximum (FWHM) values to the convergence maps.  We consider $\kappa$ values ranging from $[-5\sigma_{\kappa}, 20\sigma_{\kappa}]$, where $\sigma_{\kappa}$ is the variance of the $\kappa$ field measured from the full simulation set for each choice of smoothing filter.  The bins are linearly spaced with width $\sigma_{\kappa}/5$.  In all PDF measurements, we enforce the constraint that $\langle \kappa \rangle = 0$.

The results are plotted in Fig.~\ref{fig:opT17}. Black curves are simulation results, while red is the
analytic result.  Focusing on the upper panel (where no additional smoothing has been applied to the maps apart
from the inherent pixelization, which is treated via Eq.~\ref{eq.pixelwindow1D}), we again observe a discrepancy between the fiducial analytic result
(solid red) and the simulation.
As a further test to our hypothesis that the discrepancy can be explained with simulation resolution
effects, we again construct a $k$-space smoothing filter for the NFW profiles.
We choose the redshift-dependent filter
\begin{eqnarray}
W(k; z) &=& \frac{1}{1+k R(z)}\,;\\
R(z)    &=& 0.055\,h^{-1}\text{Mpc}\times[\log(1+z)+0.07]\,.\nonumber
\end{eqnarray}
The function $R(z)$ was chosen to give a good fit to the softening lengths employed in the T17
simulation, with some adjustments of the prefactor (our $R$ is about $10\,\%$ smaller than the
softening length).
We find the resulting one-point PDF to be largely independent of the precise functional form chosen
for $W(k; z)$, as long as it decreases steadily to $\sim 0.5$ when $kR\sim 1$.
The natural correspondence between the smoothing scale $R$ and the softening length is a further
indication that simulation resolution effects are responsible for the observed discrepancies in the
one-point PDF.

A second purpose of this section is to evaluate how well our formalism can describe the PDF of
convergence maps smoothed with a Gaussian filter.
As we explicitly show in Appendix~\ref{subapp:smoothing}, as the smoothing scale increases the PDF
becomes closer to Gaussian (as is physically expected) and receives larger contributions from the two-halo term.
These facts imply that our formalism, which is most accurate for the non-Gaussian parts of the PDF
that are dominated by massive halos, is expected to perform worse.
Thus, comparison to simulated maps is a useful test of the domain of validity.
This is plotted in the lower three panels of Fig.~\ref{fig:opT17}.
We observe that, as should be expected, the difference between solid and dashed red becomes
negligible as the smoothing scale increases.
However, our formalism does not recover the simulation PDFs very well.
As the smoothing scale increases, the PDF receives more and more contributions from non-virialized
matter, which is not included in our halo-model formalism.
One attempt to solve this problem is to convolve the analytic PDF with a Gaussian such that the
resulting variance $\langle\kappa^2\rangle$ is equal to the variance measured in the simulation.
The results of this procedure are plotted as dotted red lines.
Although the agreement (naturally) gets better, it is still far from perfect. One possible
explanation is the fact that we do not describe the negative-$\kappa$ regime accurately enough in
our formalism, and upon convolution with a relatively broad Gaussian this inaccuracy leaks into the
positive-$\kappa$ part.

Thus, we draw two conclusions from our comparison with the T17 simulations:
1) we have presented further evidence that convergence one-point PDFs measured in simulations are
susceptible to large errors due to small-scale resolution effects.
Thus, the discrepancies observed in Figs.~\ref{fig:opMassiveNuS}~and~\ref{fig:opT17} do not
invalidate our analytic formalism.
2) Our approach seems inadequate to generate accurate predictions for the PDF of convergence maps
smoothed over scales larger than a few arcminutes. Perturbative methods are likely better suited
to compute theoretical predictions in this regime.

\subsection{Halo Mass and Redshift Contributions}
\label{subsec:MZcontributions}

\begin{figure}
	\includegraphics[width=0.5\textwidth]{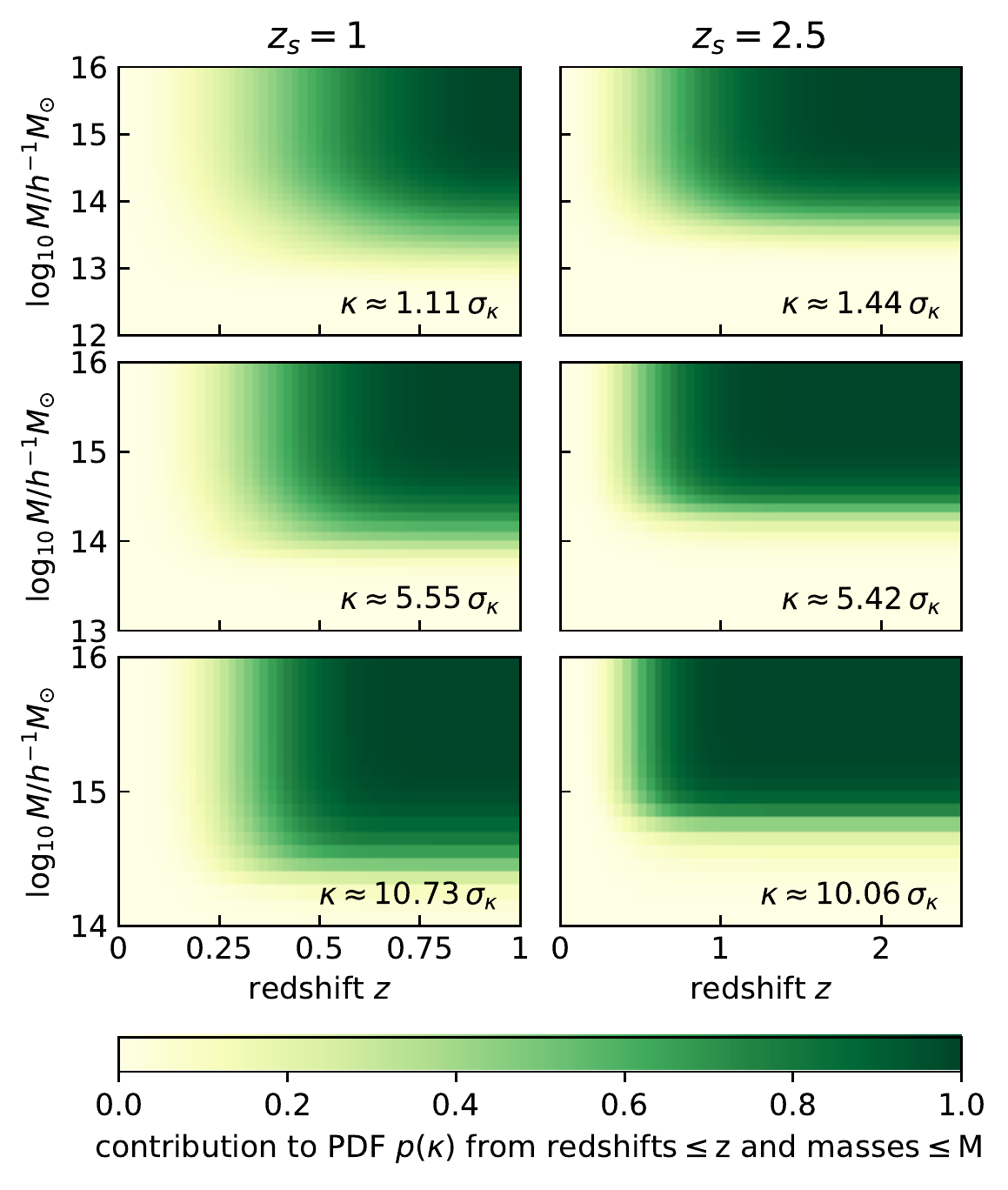}
	\caption{
	Cumulative mass and redshift contributions to the PDF for source redshifts 1 and 2.5 (columns).
	The rows represent different values of the convergence in units of the standard deviation.
	Note the different vertical scales in different rows.
	}
	\label{fig:MZcontributions}
\end{figure}

We now proceed to build up some physical intuition on the dominant contributions to the convergence
PDF. Since we label halos solely by their mass and redshift, we disentangle the contributions that
different mass and redshift intervals give to the PDF.
In Fig.~\ref{fig:MZcontributions}, we plot heat maps of the mass and redshift contributions to the
PDF for source redshifts $z_s = 1$ and $2.5$.
Each of the three rows represents a different bin of the PDF at comparable values of 
the convergence $\kappa$ in units of the respective standard deviation $\sigma_\kappa$.
Each pixel in the individual heat maps corresponds to the fraction of the final value of the
PDF if all masses and redshifts smaller than or equal to the one corresponding to the pixel are
included (thus, the upper right corner has by definition a value of one in each heatmap).
Note that overlaps make the interpretation of these heat maps somewhat complicated, in
particular for low values of $\kappa$.
In terms of redshift evolution, we can clearly see the growth of structure modulated by the
lensing kernel.
In terms of mass contributions, the intuitive picture that higher values of $\kappa$ are
sourced by more massive objects is confirmed.

\subsection{Parameter Dependence}
\label{subsec:parameterdependence}

\begin{figure}
	\includegraphics[width=0.5\textwidth]{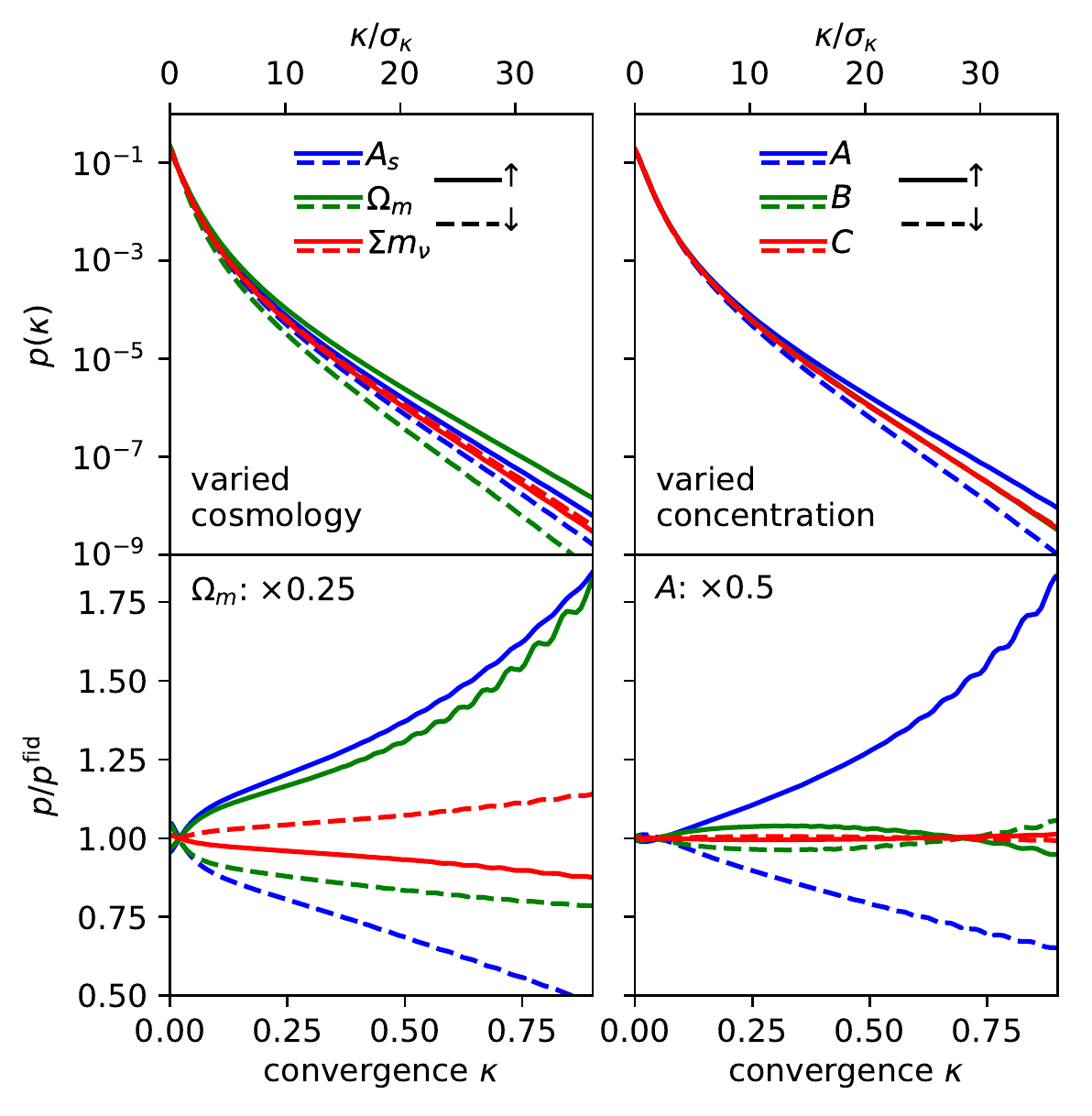}
	\caption{Dependence of the convergence one-point PDF on cosmology (\emph{left panel})
	         and concentration model (\emph{right panel}).
		 For an explanation of the parameters $A, B, C$ see the text.
		 \emph{Solid/dashed} lines represent in/de-creases of the respective parameter by
		 $10\,\%$, except for the neutrino mass sum $\Sigma m_\nu$, for which the
		 fiducial model is $0.06\,\text{eV}$ while solid/dashed represent $0.12$
		 and $0\,\text{eV}$ respectively.}
	\label{fig:varparams}
\end{figure}

In this section, we discuss the dependence of the convergence one-point PDF on the cosmological
model as well as the halo concentration-mass relation.
The results presented here are a prerequisite for the Fisher forecast in Sec.~\ref{sec:fisher},
but are also useful as a means to build up intuition.
We show results for a single source redshift $z_s = 1$.

We choose our fiducial cosmology as $h=0.7$, $\Omega_m=0.3$,
$\Omega_b=0.046$, $A_s=2.1\times 10^{-9}$, $n_s=0.97$, and $\Sigma m_\nu=0.06\,\text{eV}$.
We assume the normal hierarchy for the neutrino masses.
Following Ref.~\cite{Duffy2008}, we write the concentration-mass relation as
\begin{equation}
c(M, z) = A \left(\frac{M}{M_0}\right)^B \left(\frac{1+z}{1+z_0}\right)^C\,,
\end{equation}
where we choose $M_0 = 10^{14.5}\,h^{-1}M_\odot$, $z_0 = 0.35$, so as to break the leading
degeneracy between the three parameters $A, B, C$ (c.f. Fig.~\ref{fig:MZcontributions}).

Our results are shown in Fig.~\ref{fig:varparams}, with varied cosmology in the left panel
and varied concentration model in the right panel.
The solid/dashed lines generally represent
parameter variations by $\pm 10\,\%$, except for the neutrino mass sum, where solid corresponds to
$0.12\,\text{eV}$ and dashed to massless neutrinos.
Note that the residual curves for $\Omega_m$ and $A$ were shrunk by the stated factors to increase
readability.
With regard to varied cosmological parameters, with a fixed fractional change $\Omega_m$ has by far
the strongest influence on the PDF.
From the slight shape difference in the residual curves, we can hope that the degeneracy between
neutrino mass sum and other parameters is not too large.
With regard to the concentration model, the amplitude has by far the largest effect, while the
variation with halo mass and redshift are of minor importance.
This finding rests on the fact that the mass and redshift integrands are relatively sharply peaked
around $M_0$, $z_0$ for our simple Dirac-$\delta$ source distribution
(cf. Fig.~\ref{fig:MZcontributions}), a condition that may not be met in the case of a more
realistic source distribution.

\section{Results: Two-point PDF}
\label{sec:resultstp}

As numerical evaluation of the two-point PDF is somewhat involved,
we collect some useful simplifications in Appendix~\ref{app:numerics}.
In order to validate the analytic formalism for the two-point PDF, and by extension the covariance
matrix, we perform two tests, detailed in the following.

\subsection{Two-point correlation function}
\label{subsec:corrfunc}

\begin{figure}
\includegraphics[width=0.5\textwidth]{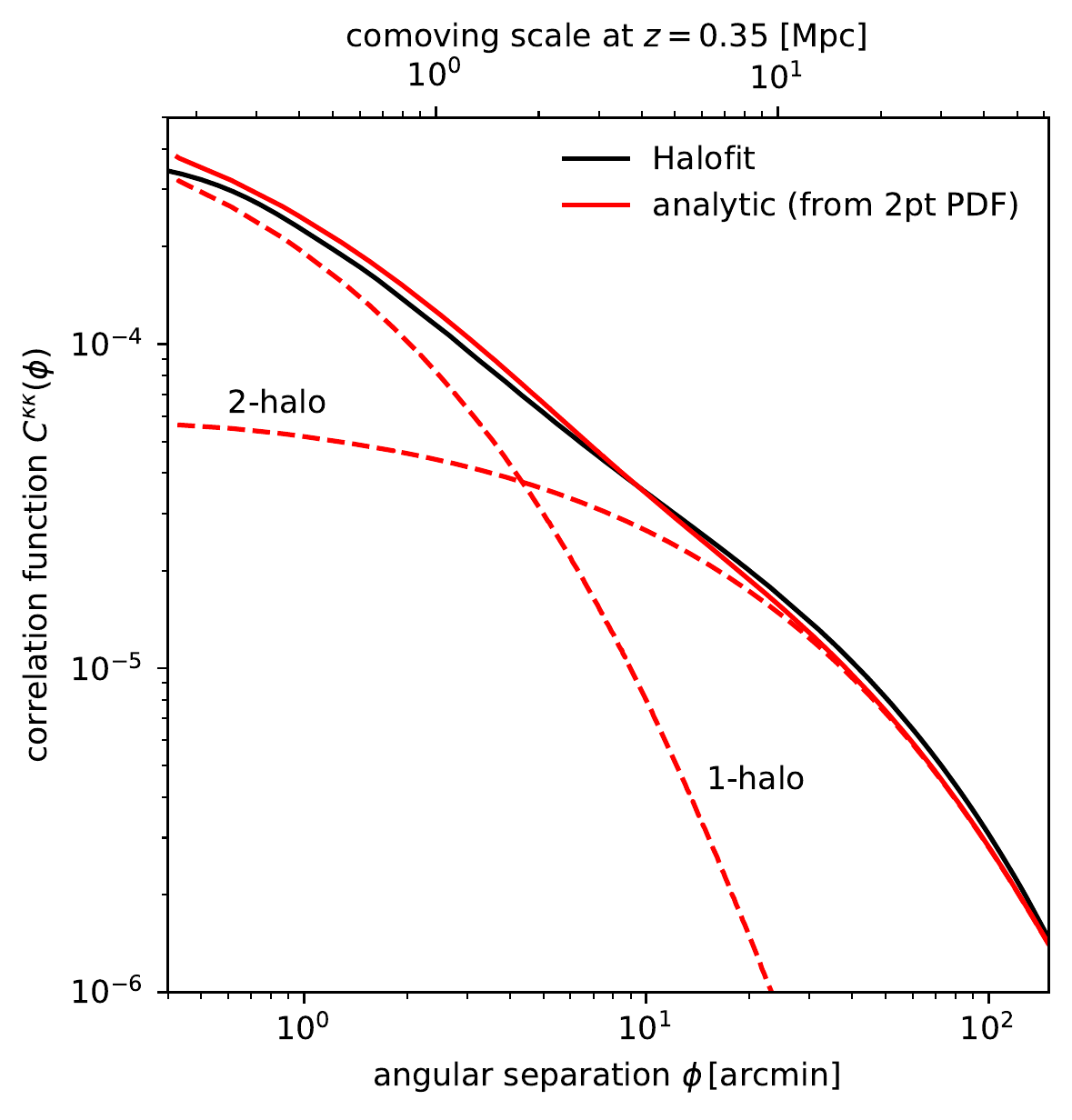}
\caption{Comparison between the 2-point correlation function reduced from our formalism for the
         2-point PDF and the Halofit computation,
	 for sources at $z_s = 1$.
	 Small discrepancies on large scales are related to the choice of radial cutoff or
	 minimum halo mass.
	 The discrepancies on small scales are likely due to inaccuracies of Halofit in the very
	 non-linear regime.}
\label{fig:corrfunc}
\end{figure}

In Fig.~\ref{fig:corrfunc}, we plot a comparison between the (isotropic) two-point correlation function obtained in our formalism with the result predicted by the Halofit fitting function
(for sources at $z_s = 1$).
Note that while the 1-halo term can be directly computed in our formalism,
the 2-halo term is simply inferred as the difference of the other two curves.
The top $x$-axis in the figure indicates the comoving scale at the approximate redshift
where the main contribution to the one-point PDF is sourced.
The small discrepancies on large scales are due to the relatively large minimum halo mass chosen in
the computation.
We confirmed that the precise choice of this cut-off has no effect on the covariance matrix.
Note that the relatively good agreement on large angular scales is a good indication that we are treating the
halo clustering effect correctly (in fact, this is the first direct validation, since the comparison
of the one-point PDF to numerical simulations is complicated by simulation resolution effects).
However, as we show in Appendix~\ref{subapp:powerspectrum},
the correlation function is not sensitive to all terms in the two-point PDF.
The discrepancies on small scales are likely stemming from the Halofit rather than the halo model
side, since we do not expect Halofit to be accurate on these rather non-linear scales. (Another possible explanation would be deviations from the NFW profile on small scales in the numerical simulations used to calibrate Halofit.)
We point out that this type of plot is useful to identify numerical instability
in a calculation of the covariance matrix.

\subsection{Covariance matrix}
\label{subsec:covmatrix}

\begin{figure}
	\includegraphics[width=0.5\textwidth]{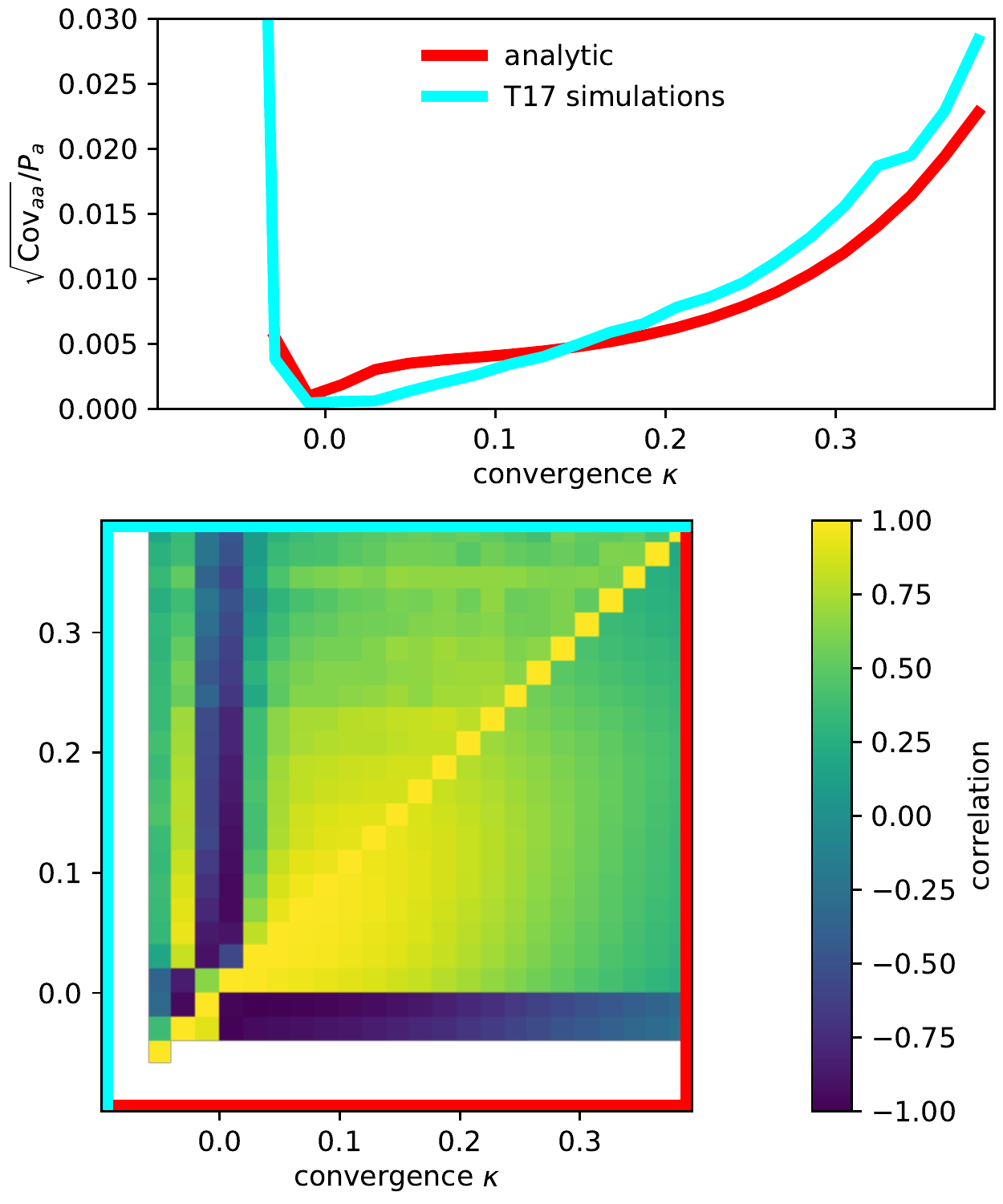}
	\caption{Comparison of the analytic and T17 covariance matrices.
	         \emph{Top panel}: diagonal elements, rescaled by the respective
		                    one-point PDFs;
		 \emph{Bottom panel}: correlation matrices (T17 in upper left triangle, analytic in lower
		                     right triangle; the color codes are as in the top panel).}
	\label{fig:covT17}
\end{figure}

Given the one- and two-point PDFs, the covariance matrix of the one-point PDF can be computed as
\begin{equation}
\text{Cov}_{ab} = \frac{1}{N_\text{pix}} \sum_\phi \left[P_{ab}(\phi) - P_a P_b\right]\,.
\end{equation}
Here, the indices label the $\kappa$-bins, and the sum runs over all pixel separations in a given
map. $N_\text{pix}$ is the number of pixels in the map, i.e. related to the sky coverage.
In practice, it is accurate enough to explicitly perform the sum over pixels for the smallest
separations and approximate the remaining summation by an integral.
Note that $P_{ab}(\phi=0) = P_a\delta_{ab}$.

We measure the covariance matrix using the 108 full-sky T17 simulations and compare it to our analytic result.
As discussed in Sec.~\ref{subsubsec:T17}, resolution issues appear to cause a discrepancy
between the analytic result and the T17 (and also MassiveNuS) simulations.
Thus, in order to make the comparison as direct as possible, we compute the covariance matrix with
the smoothed NFW profiles described before, corresponding to the dashed red line in the top
panel of Fig.~\ref{fig:opT17}.

In the bottom panel of Fig.~\ref{fig:covT17},
we show a comparison of the correlation matrices (i.e., $\text{Cov}_{ab}/\sqrt{\text{Cov}_{aa} \text{Cov}_{bb}}$).
We observe good agreement in the general structure. 
The analytic model is able to recover the transition to anti-correlation at low $\kappa$ 
(albeit displaced by about one bin).
The simulations appear to have more correlations in the high-$\kappa$ bins,
although these bins are noisy and dominated by rare events in the simulations.
There appears to be a step-like feature in the simulation covariance matrix at $\kappa\sim 0.23$,
transitioning rather suddenly from high to low correlation.
The analytic model does not predict such a feature, and since the halo model should work well at
this relatively large convergence, we are inclined to trust the model more than the simulations.

In the top panel of Fig.~\ref{fig:covT17}, we show a comparison between the diagonal elements
(we divide out the respective one-point PDFs).
While the model recovers the diagonal elements to better than an order of magnitude
(and to within $\sim 10\,\%$ for $\kappa\gtrsim 0.1$),
there are systematic differences that cannot be explained by noise in the simulations.
Given the limitations of our model with regard to non-virialized matter, we concede that the
discrepancies at low $\kappa$ are likely due to failure of the model.
However, for $\kappa\gtrsim 0.1$, we would expect the halo model to perform well, while, as we
have already seen in Sec.~\ref{subsubsec:T17}, the simulations are susceptible to resolution issues.
Although we have tried to take these into account by smoothing the NFW density profiles with a
filter calibrated on the one-point PDF,
it is likely that the covariance matrix depends differently on this filter and thus systematic
discrepancies are still to be expected.  This also highlights the possible dangers in purely simulation-based inference procedures, which could be biased by such resolution effects.

\section{Fisher forecast}
\label{sec:fisher}

In this section, we present a simple Fisher matrix parameter forecast using the WL convergence PDF.
We assume a Rubin-Observatory-like survey with sky coverage $20,000\,\text{deg}^2$, pixel size
$\Omega_\text{pix} = (0.41\,\text{arcmin})^2$ and source density $n_\text{gal}=45\,\text{arcmin}^{-2}$.
The latter number is taken from Ref.~\cite{Liu2018WLPDFneutrinomass};
in contrast to this work we make the simplifying assumption of a Dirac-$\delta$ source distribution
at $z_s = 1$.
(Note that the inclusion of tomographic information would further improve the forecast error bars computed here.)
Our fiducial cosmological model is the same as in Sec.~\ref{subsec:parameterdependence}.

In order to upweight the cosmological signal with respect to the shape noise,
we convolve the convergence profiles with a Wiener filter, constructed as
\begin{equation}
F_\ell = C^{\kappa\kappa}_\ell/(C^{\kappa\kappa}_\ell + N_\ell)\,,
\end{equation}
where we compute the convergence power spectrum using Halofit and
the noise power spectrum $N_\ell = 0.3^2/n_\text{gal}$ is flat.
Applying the Wiener filter is crucial in optimizing the sensitivity of the one-point PDF,
since it is a real-space statistic.

The shape noise in the filtered maps has the correlation function
\begin{equation}
\label{eq.shapenoisecorr}
\zeta_{\kappa,\text{sn}}(\phi) = \sum_\ell\frac{\ell+1/2}{2\pi}
	N_\ell F_\ell^2 W_{\text{2pt},\ell}^\text{pix} P_\ell(\cos\phi)\,,
\end{equation}
where $W_{\text{2pt},\ell}^\text{pix}$ is the pixel window function,
\begin{equation}
W_{\text{2pt},\ell}^\text{pix} = \frac{4}{\pi}\int_0^{\pi/4} d\varphi\,
         \left[\text{sinc}(\cos(\varphi)\ell a)\text{sinc}(\sin(\varphi)\ell a)\right]^2\,,
\end{equation}
where $a$ is half the pixel side length.
The ``noisy" one-point PDF and covariance matrix are computed as
\begin{eqnarray}
p_{\text{sn},a} &=& (\mathcal{G}_1[P_i])_a\\
\text{Cov}_{\text{sn},ab} &=&  \sum_\phi \big[ (\mathcal{G}_2^{(\phi)}[P_{ij}(\phi)])_{ab}\nonumber\\
&&\hspace{2em} -(\mathcal{G}_1[P_i])_a (\mathcal{G}_1[P_j])_b \big]\,,
\end{eqnarray}
where $\mathcal{G}_n[\cdot]$ indicates convolution with an $n$-dimensional Gaussian.
Here, $\mathcal{G}_1$ has variance $\zeta_{\kappa,\text{sn}}(0)$
and $\mathcal{G}_2^{(\phi)}$ has the covariance matrix
\begin{equation}
C_\text{sn}^{(\phi)} = 
\begin{pmatrix}
\zeta_{\kappa,\text{sn}}(0)    & \zeta_{\kappa,\text{sn}}(\phi) \\
\zeta_{\kappa,\text{sn}}(\phi) & \zeta_{\kappa,\text{sn}}(0)
\end{pmatrix}\,.
\end{equation}

\begin{figure}
	\includegraphics[width=0.5\textwidth]{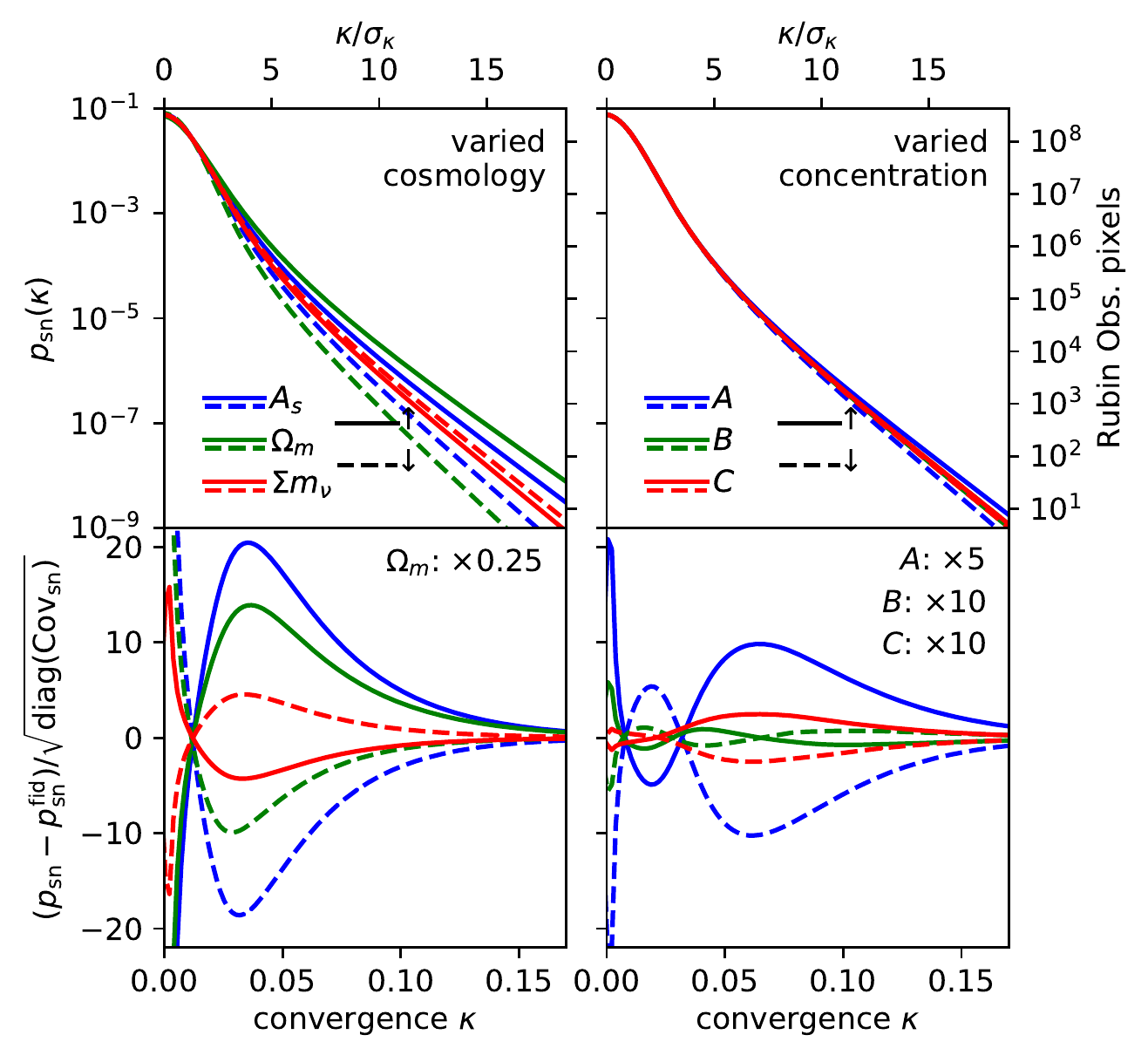}
	\caption{The convergence one-point PDFs, as a function of parameter variations,
	         including shape noise of $\sigma_{\kappa,\text{sn}} = 0.2$.
		 The plotting conventions are the same as in Fig.~\ref{fig:varparams},
		 however, since in this plot we applied a Wiener filter to the convergence profiles
		 the convergence values on the $\kappa$-axis are not directly comparable.
		 As in Fig.~\ref{fig:varparams}, all parameter variations are $10\,\%$, except for
		 $\Sigma m_\nu$ for which it is $100\,\%$.
		 The bottom row shows the ratio of the residuals to the square-root of the diagonal
		 elements of the covariance matrix.}
	\label{fig:varparamssn}
\end{figure}

In Fig.~\ref{fig:varparamssn}, we plot the same PDFs already shown in Fig.~\ref{fig:varparams},
this time including the shape noise contribution.
We bin the PDFs into bins of width $\Delta\kappa = 1.65\times 10^{-3}$,
and quote the number of pixels expected in the model survey.
Again, the parameter variations are $10\,\%$ for all parameters except the neutrino mass sum,
for which the variation is $100\,\%$ (dashed is massless neutrinos, solid is $\Sigma
m_\nu=0.12\,\text{eV}$).
Note that, for clarity, we rescaled some of the curves in the bottom row of Fig.~\ref{fig:varparamssn}
by the given amounts.
It is interesting to see that the Wiener filter not only serves the purpose of minimizing the noise
contribution, but also increases the sensitivity of the PDF on cosmology in comparison to the
concentration model (compare Fig.~\ref{fig:varparams}, which has the same parameter variations).
This can be understood as a consequence of the suppression of small-scale power
which makes the exact shape of the convergence profiles (and thus the concentration model)
less relevant, while the main dependence on the cosmological model comes through the halo mass
function.
This will be crucial for the parameter forecast.

We compute the Fisher matrix as
\begin{eqnarray}
F_{ab} &=& \frac{\partial\myvector{p}^T}{\partial\theta^a}\text{Cov}^{-1}
         \frac{\partial\myvector{p}}{\partial\theta^b} \nonumber\\
	 &+& \frac{1}{2}\text{tr}\,\text{Cov}^{-1}\frac{\partial\text{Cov}}{\partial\theta^a}
	                         \text{Cov}^{-1}\frac{\partial\text{Cov}}{\partial\theta^b}\,,
\end{eqnarray}
where $\myvector{p}$ is the binned one-point PDF and $\theta$ is the parameter vector (indexed by $a,b$).
Contrary to conventional wisdom, the second term in the Fisher matrix is not always negligible and
changes parameter constraints by a few $10\,\%$.
In the data vector, we include the PDF in the interval $\kappa\in[-0.03, 0.17]$;
we find the constraints to be relatively independent of the choice of binning.
It is worth noting that we are including some negative-$\kappa$ part of the PDF,
which is not well described by our model.
However, due to the Wiener filter and noise convolution it is rather challenging to cleanly exclude this regime,
and as we will argue below we do not believe that keeping the uncertain values invalidates
the major conclusions from our forecast.

We consider six free parameters, namely $\{A_s, \Omega_m, \Sigma m_\nu\}$ on the cosmology side
and $\{A, B, C\}$ in the parametrization of the concentration model from Ref.~\cite{Duffy2008}.
For the concentration model, we choose the mass and redshift pivots described at the end of
Sec.~\ref{subsec:parameterdependence} (in the end we rescale $A$ to the original value).

\begin{SCfigure*}
	\includegraphics[width=0.7\textwidth]{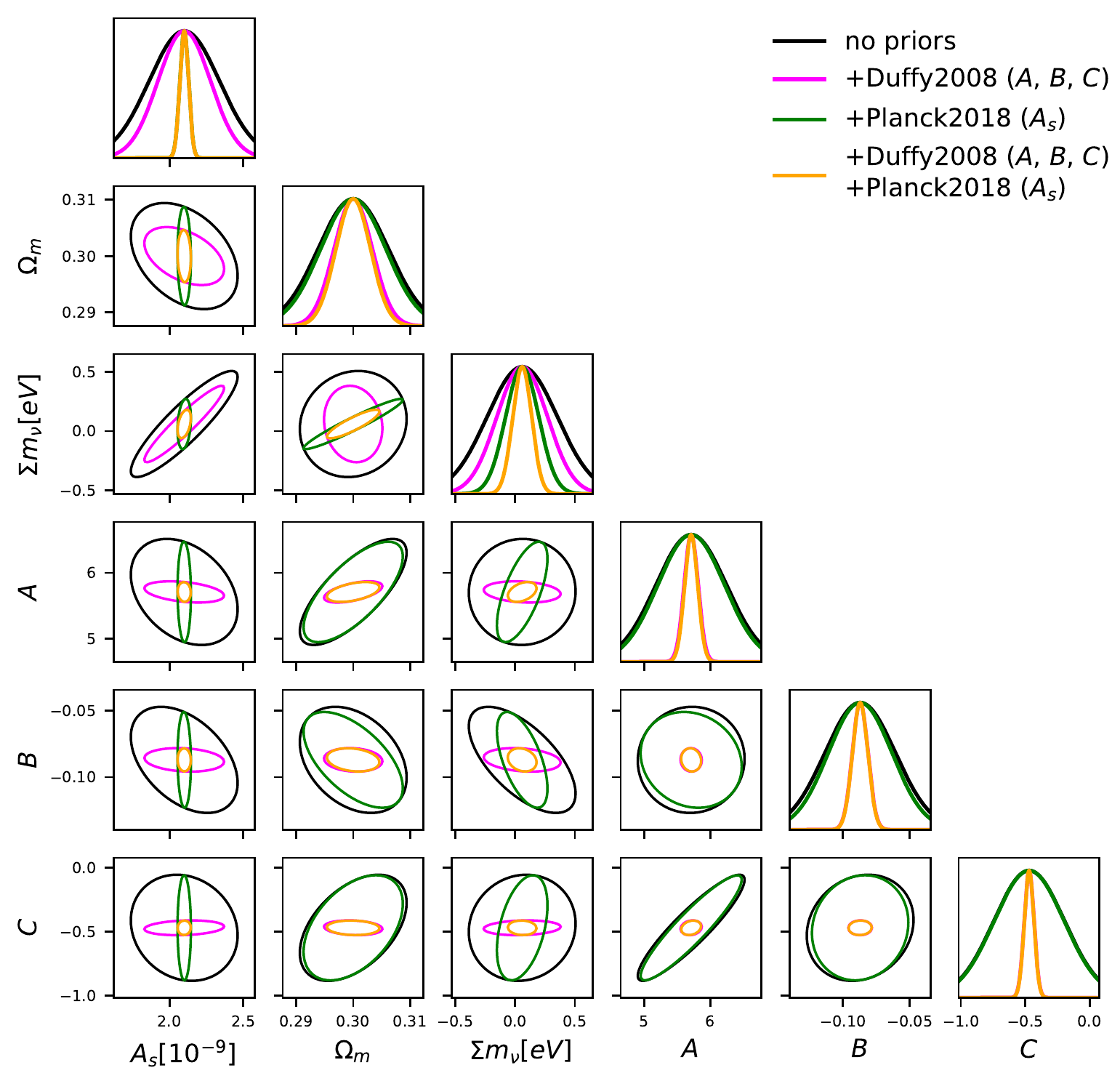}
	\caption{Four Fisher forecasts for parameter constraints
	         from the weak lensing one-point PDF.
	         We are assuming Gaussian shape noise as described in the text,
	         Rubin Observatory sky coverage of $20,000\,\text{deg}^2$,
		 and a source density of $45\,\text{arcmin}^{-2}$ in a Dirac-$\delta$ distribution
		 at redshift $z_s=1$.
		 The ellipses are $1\sigma$ confidence intervals.
		 \emph{Black}: both cosmology and concentration model are completely
		               inferred from the data;
		 \emph{magenta}: including simulation priors from Ref.~\cite{Duffy2008}
		                 on the concentration model;
		 \emph{green}: including CMB prior from Ref.~\cite{Planck2018} on $A_s$;
		 \emph{orange}: including both priors on the concentration model and $A_s$.}
	\label{fig:fisher}
\end{SCfigure*}

\begin{table}
	\begin{tabular}{lccc|ccc}
	                              & \multicolumn{6}{c}{$\sigma(X)/X_\text{fid}\,[\%]$} \\
	                              & $A_s$ & $\Omega_m$ & $\Sigma m_\nu$ & $A$  & $B$ & $C$ \\
	\hline
	no priors                     & 11    & 2.1        & 490            & 9.3  & 30  & 58  \\
	+Duffy2008                    & 8.5   & 1.1        & 350            & 1.9  & 6.7 & 8.1 \\
	+{\it Planck}2018 ($A_s$)      & 1.4   & 1.9        & 230            & 8.8  & 27  & 58  \\
	+Duffy2008+{\it Planck}2018 ($A_s$) & 1.4   & 1.0        & 132            & 1.7  & 6.6 & 8.0 \\
	\end{tabular}
	\caption{$1\sigma$ constraints on cosmological and concentration model parameters,
	         for four different choices of priors.
		 The numbers are relative to the fiducial value, in percent.}
	\label{tab:sigmas}
\end{table}

In Fig.~\ref{fig:fisher}, we plot four different Fisher forecasts,
differing solely by the priors we place on the concentration model
and the scalar fluctuation amplitude $A_s$.
Black includes no priors at all,
magenta includes the error bars on $A$, $B$, $C$ from Ref.~\cite{Duffy2008}
as diagonal Gaussian priors,
green includes the CMB prior from Ref.~\cite{Planck2018} on $A_s$,
and orange includes both priors.
For clarity, in Table~\ref{tab:sigmas} we quote the fractional $1\sigma$ constraints (in percent) on the three cosmological parameters 
as well as the concentration model parametrization for the different choices of priors.
We observe that if priors on the concentration model as well as $A_s$ are included,
our results suggest that the WL convergence PDF alone can provide an error bar on $\Sigma m_{\nu}$ comparable to the minimum possible neutrino mass sum. Including tomographic information and/or the full WL convergence power spectrum would only improve these constraints.

Our simple Fisher forecast has several limitations:
\begin{itemize}
	\item We are assuming a Gaussian likelihood even though we know that the one-point PDF
	      is a non-Gaussian statistic.  Unfortunately, the full likelihood for this observable
	      is not yet known.  Evidence of a small bias when assuming a Gaussian likelihood was
	      seen for the tSZ PDF in Ref.~\cite{Hill2014}.  However, a Gaussian likelihood was used
	      and shown to be unbiased for the WL PDF in Ref.~\cite{Liu2018WLPDFneutrinomass},
	      albeit with the caveat that high-$\kappa$ bins were removed, which Gaussianizes the
	      statistic. The PDF observable could be a useful opportunity to apply new methods in
	      likelihood-free inference~(e.g., Refs.~\cite{Alsing2018,Alsing2019}).
	\item We work in the Fisher approximation; however, MCMC results from Ref.~\cite{Liu2018WLPDFneutrinomass} indicate that this is not a bad approximation.
	\item Our analytic covariance matrix is likely not exact for small values of $\kappa$;
	      however, our results from Sec.~\ref{subsec:covmatrix} indicate that the formalism
	      tends to overestimate the covariance matrix in this regime, which implies that in
	      this respect our forecast parameter constraints are conservative.
	\item Our formalism is unable to make exact predictions for the
	      negative-$\kappa$ tail. However, this loss of detail also is likely to imply
	      that our forecast is conservative.  Including full information from the negative-$\kappa$ region would only improve the constraints found here.
	\item We make the simplifying assumption of a Dirac-$\delta$ source distribution in a single
	      bin.
	      Accounting for a realistic spread in the source distribution would likely not significantly affect the constraints; moreover,
	      Ref.~\cite{Liu2018WLPDFneutrinomass} has demonstrated
	      that tomography with multiple source bins has the potential to significantly improve
	      the constraints.
	\item It is not clear that the Wiener filter employed is optimal.
	      One possibility would be to consider the PDFs of maps smoothed on various scales
	      simultaneously.  This approach should recover some of the scale-dependent information contained in the full $N$-point functions that is lost when compressing to the zero-lag one-point PDF.
	      The cross-covariance between the PDFs should be easy to compute in a simple extension
	      of the formalism presented above (all terms in our model for the covariance matrix are
	      symmetric under interchange of two convergence profiles, the modification should amount to
	      breaking that symmetry by filtering them with two different kernels).
	\item We do not consider any cosmology dependence of the concentration model (although this is likely small),
	      and our prior on the concentration model parameters may be optimistic,
	      particularly given baryonic effects on the small-scale matter distribution.
\end{itemize}
The last point implies that a more robust understanding of the concentration-mass relation and halo density profiles in general would be extremely beneficial for parameter inference from the WL PDF, similar to the WL power spectrum.

Our forecast is similar in set-up to the simulation-based one presented in Ref.~\cite{Liu2018WLPDFneutrinomass}.
However, there are a few key differences:
(1) they do not make the simplifying assumption of a Dirac-$\delta$ source distribution;
(2) they filter the convergence maps with a different $\ell$-space filter;
(3) they are able to use more of the negative-$\kappa$ regime;
(4) they (naturally) cannot marginalize over small-scale modeling uncertainties;
(5) they do not include noise correlations, which we find to have an appreciable effect.
As a consequence of these differences, we find the agreement between our results and theirs to be
satisfactory. For a rough qualitative comparison, we consider our result with the
concentration-model prior (i.e. the second row in Table~\ref{tab:sigmas}),
where we have $\sigma(10^9 A_s)=0.18$, 
	    $\sigma(\Omega_m)=0.003$,
	    $\sigma(\Sigma m_\nu)=0.21\,\text{eV}$.
We find that including the effect of noise correlations gives about a factor
of 2 degradation in constraints, which is approximately offset by increasing
the source number density by a factor 4.
Thus, for the purposes of this rough comparison, we choose the turquoise
contours in Fig.~3 of Ref.~\cite{Liu2018WLPDFneutrinomass} as reference
(these do not include noise correlations and have a source density of
$13.25\,\text{arcmin}^{-2}$).
Approximating the posterior as Gaussian,
we read off $\sigma(10^9 A_s)=0.1$,
	  $\sigma(\Omega_m)=0.006$,
	  $\sigma(\Sigma m_\nu)=0.12\,\text{eV}$.
Thus, our halo-model-only forecast reproduces these simulation-derived
constraints to within a factor of two.
Note, however, the difference in orientation of ellipses involving
$\Omega_m$.

\begin{figure}
	\includegraphics[width=0.5\textwidth]{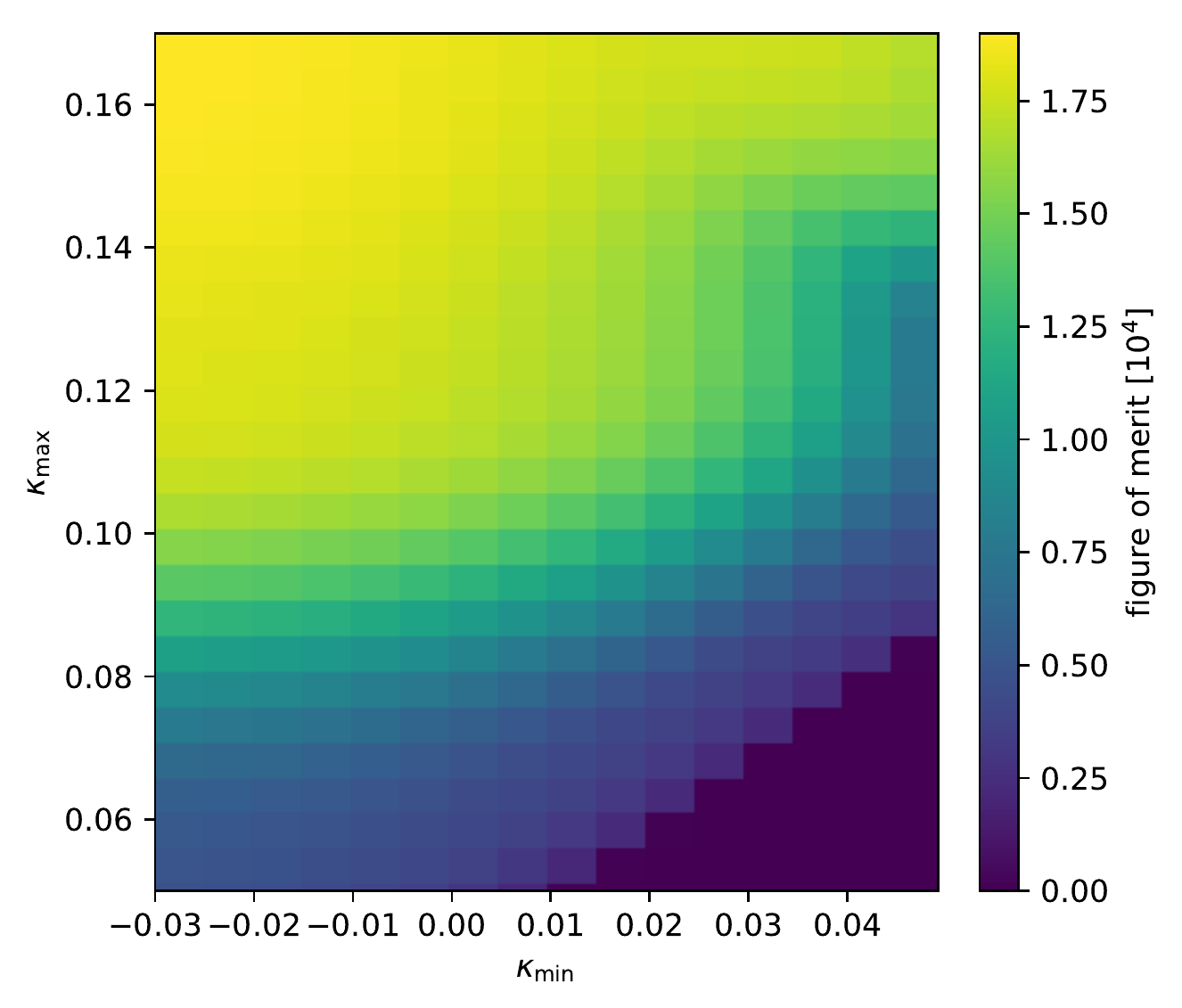}
	\caption{Dependence of the normalized 3-parameter figure of merit (FOM),
	         as defined in Eq.~\eqref{eq:fom},
		 on the minimum/maximum $\kappa$-cutoff.
		 We observe that the negative-$\kappa$ part of the PDF contains substantial
		 information, and that our fiducial choice of $\kappa_\text{max} = 0.17$ extracts essentially all of the information content (assuming $\kappa_\text{min} = -0.03$).}
	\label{fig:fom}
\end{figure}

As a final result of this section, we explore the constraints' dependence on the choice of the minimum
and maximum convergence values in the data vector $\myvector{p}$.
We consider the 3-parameter figure of merit,
\begin{equation}
\text{FOM} = \left(\text{det} F_\text{cosmo}^{-1}\right)^{-1/3}\,,
\label{eq:fom}
\end{equation}
where $F_\text{cosmo}$ is the sub-block of the Fisher matrix corresponding to the cosmological
parameters $\{A_s,\Omega_m,\Sigma m_\nu\}$.
We work with the maximum set of priors, corresponding to the orange lines in Fig.~\ref{fig:fisher}. 
A heat map of this quantity as a function of minimum and maximum cut-off is shown in
Fig.~\ref{fig:fom}.
Again, we emphasize that the ``convergence" values quoted there are related through the Wiener filter
to the physical convergence, making the interpretation somewhat difficult.
The first conclusion from Fig.~\ref{fig:fom} is that the minimum cut-off $\kappa_\text{min}$ is
rather important for the constraining power.
Second, for the minimum cut-off chosen in our analysis above, $\kappa_\text{min} = -0.03$,
the information content of the positive-$\kappa$ tail is essentially saturated
with our choice of $\kappa_\text{max} = 0.17$.  In fact, using a somewhat smaller value of $\kappa_\text{max}$ could serve to Gaussianize the likelihood for the WL PDF, evidently without a significant loss in constraining power.

\section{Conclusions}
\label{sec:conclusions}

We have developed a halo-model-based formalism for the weak lensing convergence one-point and
two-point PDFs (and, by extension, the covariance matrix of the one-point PDF).
The strengths of our model lie in its superior speed compared to simulations,
the ability to explicitly marginalize over small-scale uncertainties (parametrized through the
concentration model), and its interpretability.

As expected on physical grounds, the accuracy of our model is highest in the positive-convergence
tail.
We have shown that, in this regime, discrepancies in the one-point PDF with respect to numerical simulations
are explained by simulation resolution issues and do not invalidate our method.
It may be argued that as soon as the convergence map is smoothed on a sufficient scale, the
simulation resolution effects would be less severe.
However, at the wave numbers at which the MassiveNuS simulations start to show appreciable power
deficiencies, the Wiener filter employed in this work still assumes values of order 0.1;
thus, smoothing on a reasonable scale appears not sufficient to neutralize the small-scale issues in
the simulations.

On the other hand, in the negative convergence regime and for large smoothing scales our formalism
is less accurate.
Alternative approaches are likely better suited for accurate theoretical predictions in these
regimes.

Validation of the covariance matrix was found to be challenging,
and discrepancies with respect to the simulations remain over a range of convergence values.
However, discrepancies at low $\kappa$ are likely irrelevant in any real analysis
due to the dominance of the shape noise contributions there.
The smaller discrepancies at high $\kappa$ could simply be
due to resolution effects in the simulations,
as we already demonstrated for the one-point PDF itself.

Using our formalism, we have performed a Fisher forecast in the
$\{A_s, \Omega_m, \Sigma m_\nu\}$ parameter space for a Rubin-Observatory-like survey.
We have found that the convergence PDF alone could provide a $1\sigma$ error bar on $\Sigma m_{\nu}$
that is comparable to the minimum neutrino
mass sum allowed from oscillation experiments,
if a CMB prior on $A_s$ and simulation priors on the concentration-mass relation are included.
Our results are in good agreement with previous simulation-based forecasts,
and could be generated in a fraction of the time.
We have also presented arguments why the several limitations of our simple forecast
are likely to render it conservative, i.e., a more sophisticated analysis would probably
find a further improvement in constraints (although this gain could be negated by the inclusion and marginalization
of systematic errors in the measurement of convergence values).
Our work demonstrates that an analytic approach to non-Gaussian WL statistics is feasible for upcoming surveys,
at least in terms of the statistical constraining power.
Tests for biases will necessitate end-to-end simulation analyses, which are beyond the scope of this work.

We believe that a comprehensive model for the weak lensing convergence one-point PDF and its
covariance matrix would be most accurate if it combines different approaches.
For example, one could imagine taking simulation results for the negative-convergence and Gaussian
part of the PDF, while the positive-convergence tail is generated with our formalism.
A desirable side effect of this method could be that smaller simulation volumes are required in
order to sample the quasi-linear part of the density field.

Possible extensions of our model could involve the use of compensated density profiles
(e.g., Ref.~\cite{ChenAfshordi2020}
which could help improve accuracy near the Gaussian peak of the PDF),
the effective halo model approach from Ref.~\cite{Philcox2020},
and the inclusion of voids.

\acknowledgments

We would like to thank Peter Melchior, David N. Spergel, Zoltan Haiman, Jia Liu, and Cora Uhlemann for useful discussions.
Research at Perimeter Institute is supported in part by the Government of Canada
through the Department of Innovation, Science and Economic Development Canada
and by the Province of Ontario through the Ministry of Economic Development, Job Creation and Trade.  JCH thanks the Simons Foundation for support.
KMS was supported by an NSERC Discovery Grant, an Ontario Early Researcher Award, a CIFAR fellowship,
and by the Centre for the Universe at Perimeter Institute.



\appendix

\section{Some analytic calculations}
\label{app:analytic}

\subsection{Proof that we recover the power spectrum}
\label{subapp:powerspectrum}

We can show quite easily that the 2-point function is given by
\begin{equation}
C(\phi) = \left[\partial^2_{\ii\jj} \log P_{\ii\jj}\right]_{\lambda_\ii=\lambda_\jj=0}\,,
\end{equation}
where $\partial_a\equiv -i\partial/\partial\lambda_a$.
Note that the 1-point factors in the 2-point PDF give no contribution, since their
logarithm is a sum of functions that depend only on $\lambda_\ii$ or $\lambda_\jj$.
Since the Fourier-space PDF $P_{\ii\jj}$ is a product of one- and two-halo term,
the correlation function becomes a sum.

The 1-halo term is given by Eq.~\eqref{eq:twoPDFonehalo}:
\begin{equation}
\log P^\text{1h}_{\ii\jj}\supset \int_{M,z,\ell} K_\ii^{(\ell)}K_\jj^{(\ell)}J_0(\ell\phi)\,.
\end{equation}
Performing the differentiation, we obtain
\begin{equation}
C^\text{1h}(\phi) = \int_\ell J_0(\ell\phi) \int_{M,z}
	\left[\int_\theta \kprof(\theta) J_0(\ell\theta)\right]^2\,,
\end{equation}
which is equivalent to the standard expression
\begin{equation}
C^\text{1h}(\ell) = \int_{M,z} |\tilde{\kappa}_\ell|^2\,.
\end{equation}

The 2-halo term is given by Eq.~\eqref{eq:twoPDFtwohalo}:
\begin{eqnarray}
\log P^\text{2h}_{\ii\jj}&\supset&\int_z H \big[\alpha_\ii\alpha_\jj\zeta(\phi)
                          +\frac{1}{2}\beta_{\ii\jj}^2\zeta(0) \nn\\
&&\hspace{1.0cm}
+\beta_{\ii\jj}(\alpha_\ii+\alpha_\jj)\zeta\big(\frac{\phi}{2}\big)\big]\,.
\end{eqnarray}
Since $\alpha_\ii = 0$, $\beta_{\ii\jj} = \partial_\ii\beta_{\ii\jj} = 0$
when $\lambda_\ii = \lambda_\jj = 0$, only the first term in the square brackets survives, and we get
\begin{equation}
C^\text{2h}(\phi) = \int_z H\zeta(\phi)\left[\int_{M,\theta} b \kprof(\theta)\right]^2\,.
\end{equation}
Transforming to conjugate space, this gives
\begin{equation}
C^\text{2h}(\ell) = \int_z\frac{H}{\chi^2}P_\text{lin}(\ell/\chi, z)\left[\int_{M,\theta} b
\kprof(\theta)\right]^2\,.
\label{eq:powerspectrumfromtwopoint}
\end{equation}
We see that this is not exactly equal to the usual halo model calculation,
which would have an extra Bessel function in the square brackets.
The difference, as remarked above, arises from the fact that we approximate the
linear overdensity field as approximately constant over the typical size of a halo,
so that the linear power spectrum becomes negligible whenever the argument $\ell\theta$
of the neglected Bessel function would become appreciable.

\subsection{Large smoothing limit}
\label{subapp:smoothing}

In this section we briefly discuss how smoothing the convergence field with a Gaussian filter of
large aperture affects the 1-point PDF.
Denoting the smoothing scale by $\sigma$, we have for the smoothed convergence profiles
\begin{equation}
\kappa_\sigma(\theta) = {\rm Gaussian}_\sigma(\theta) * \kprof(\theta)\,,
\end{equation}
which, after inserting the conjugate space expressions, leads to
\begin{equation}
\kappa_\sigma(\theta) = \int_{\theta',\ell} \kprof(\theta')
        J_0(\ell\theta)J_0(\ell\theta')e^{-\ell^2 \sigma^2/2}\,.
\end{equation}
In general, the ratio $\theta'/\sigma$ will attain its maximum when $\theta'$ is comparable
to the projected halo radius. Of course, this varies with halo mass and redshift,
but it is still reasonable to formally introduce a scale $\hat\theta$ that characterizes
typical halo extents.
Since we assume $\sigma$ to be large, we can expand in $\hat\theta/\sigma$.

The zero-order term in the expansion parameter $\hat\theta/\sigma$ is given by
\begin{equation}
\kappa^{(0)}_\sigma(\theta) = \sigma^{-2}e^{-\theta^2/2\sigma^2}\bar\kprof\,,
\end{equation}
where we have introduced
\begin{equation}
\bar\kprof \equiv \int_\theta \kprof(\theta)\,.
\end{equation}
It simply measures the total amount of signal in a single halo, which is then smoothed into a
Gaussian by the $\sigma$-filter.

We now compute the $n$-th order cumulants $k_n$. Similarly to the power-spectrum calculation
performed in the previous section, the cumulants are related to derivatives of $\log P$,
so that the cumulants naturally split into one- and two-halo terms:
\begin{eqnarray}
k_n^\text{1h} &=& \left[ \partial^n_\ii \log P^\text{1h}_\ii \right]_{\lambda_\ii=0} \nn \\
&=& \frac{2\pi}{n} \sigma^{2-2n} \int_{M,z} \bar\kprof^n\,; \\
k_n^\text{2h} &=& \left[ \partial^n_\ii \log P^\text{2h}_\ii \right]_{\lambda_\ii=0}
\nn \\
&=& \sigma^{4-2n} \int_z \frac{H\zeta(0)}{2} \sum_{m=1}^{n-1} {n\choose m}
      \frac{2\pi}{m} \frac{2\pi}{n-m} \nn\\
&\times& \left[\int_M b \bar{\kprof}^{n-m}\right]
         \left[\int_M b\bar{\kprof}^m   \right]\,.
\end{eqnarray}
Thus, we see that the scalings with the expansion parameter are
\begin{equation}
k_n^\text{1h} \propto (\hat\theta/\sigma)^{2n-2}\,;\quad
k_n^\text{2h} \propto (\hat\theta/\sigma)^{2n-4}\,.
\end{equation}
From this, we can draw two conclusions:
\begin{enumerate}
\item Increasing the order of a cumulant by one introduces two powers of the expansion parameter.
Thus, in the limit where the expansion parameter is small, we converge at a Gaussian distribution.
\item The cumulants arising from the clustering term are larger by two powers of the expansion
parameter, so that for large enough smoothing scales the two-halo term eventually takes over
(despite being only a relatively small correction in the unsmoothed case).
\end{enumerate}

\section{Numerical evaluation}
\label{app:numerics}

For efficient computation of the one- and especially the two-point PDF,
two observations can be made.
First, integrals of the form
\begin{equation}
\int_{M,\theta} e^{i\kprof(\theta)\lambda}
\end{equation}
can be transformed to
\begin{equation}
\int d\kappa\,e^{i\kappa\lambda} \int_M \pi\frac{d\tprof^2(\kappa)}{d\kappa}\,,
\end{equation}
where $\tprof(\kappa)$ is the inverse function of the convergence profile $\kprof(\theta)$.
Thus, we can use the FFT.
Note that this requires that the convergence profiles are invertible (monotonic).
Application of $\ell$-space filters (such as the pixel window function or the Wiener filter)
can occasionally lead to non-monotonic profiles;
in that case we split the integral into segments in which the convergence profiles
are monotonic.

Second, in the two-point PDF it is not necessary to compute the $K_\ii^{(\ell)}$
from Eq.~\eqref{eq:Kiell} explicitly,
since the integral over $\ell$ reduces to
\begin{equation}
\int_\ell J_0(\ell\theta_\ii)J_0(\ell\theta_\jj)J_0(\ell\phi)\,,
\end{equation}
which has the analytic form \citep[pg.~411~Eq.~(3)]{WatsonBessel}
\begin{equation}
[ 4\pi^2\Delta(\theta_\ii, \theta_\jj, \phi) ]^{-1}\,.
\end{equation}
Here, $\Delta$ denotes the area of the triangle with the arguments as its sides.
If no triangle can be formed, the integral vanishes.
The case $\phi=0$ is relevant for this work; then the integral
is a multiple of $\delta_D(\theta_\ii-\theta_\jj)$ and this propagates through in such
a way that the zero-separation two-point PDF simplifies to
\begin{equation}
P(\kappa_\ii,\kappa_\jj;\phi=0) = P(\kappa_\ii)\delta_D(\kappa_\ii-\kappa_\jj)\,,
\end{equation}
as it should.

\section{Validating Assumptions of the Halo Model Approach}
\label{app:assumptions}

There are two distinct classes of assumptions made in this work.
The first class comprises the basic underpinnings of the halo model.
In comparing to two different sets of simulations, we have seen that we can describe the convergence
PDF accurately well into the non-Gaussian tail, while our model has deficiencies for $\kappa
\lesssim 0$ and for large smoothing scales. These problems are consistent with intuitive expectation
-- we know that any halo model formalism would face these problems.
On the other hand, the second class of assumptions refers to certain technical choices made in the
formalism that could in principle be dropped without leaving the realm of the halo model.

\subsection{Born approximation}
\label{subapp:born}

The first of these is the Born approximation. Our formalism crucially requires this approximation,
since it relies on the additivity of the convergence signal.
While we do not examine the Born approximation in any detail here, we note two reasons why we believe
it to constitute only a minor correction to the PDF:
(1) we performed a simple numerical test in which we placed a single halo of mass
$10^{14}\,h^{-1}M_\odot$ at $z = 0.5$ and ray traced a parallel beam through its potential.
The resulting deflection angles are in extremely good agreement with the Born approximation.
Although this does not test for higher-order effects such as lens-lens coupling, it still
constitutes a simple test demonstrating the smallness of post-Born terms
(2) some tests have already been performed in the literature
\cite{PetriHaimanMay2017, Barthelemy2020},
seemingly coming to the conclusion that post-Born terms are quite negligible for galaxy lensing.

\subsection{Triaxialiaty}
\label{subapp:triaxiality}

\begin{figure}
	\includegraphics[width=0.5\textwidth]{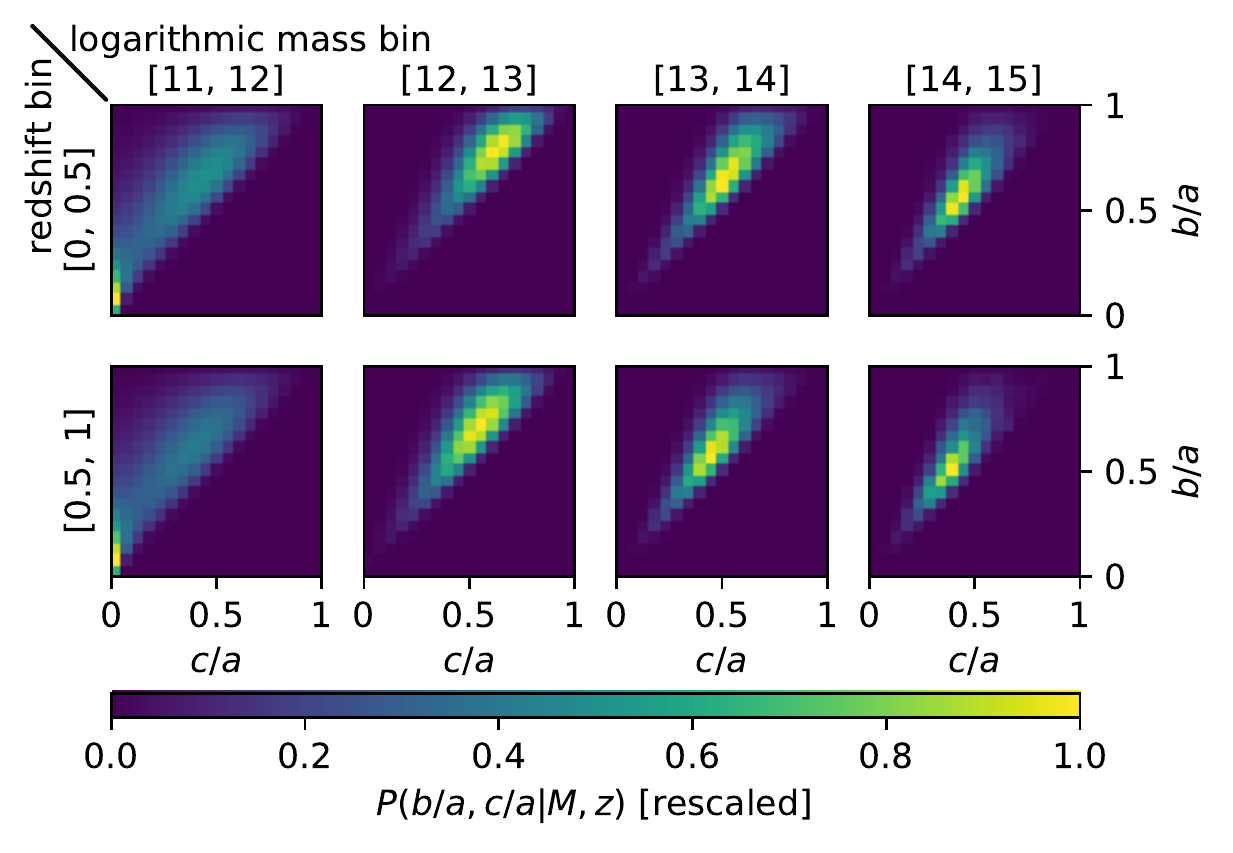}
	\caption{
	Measured axes ratio distribution from the MassiveNuS halo catolog,
	used to construct the simplified simulations that we utilize to
	assess the effect of ellipsoidal vs spherical halos
	(cf. Fig.~\ref{fig:triaxiality}).
	Note that the lower triangular part of these matrices is necessarily
	zero by the definition of $a>b>c$.
	We note the peaks near $b=c=0$ in the lowest mass bin,
	which are due to imprecise axes ratio measurements and spurious halo
	identifications close to the simulation's resolution limit.
	}
	\label{fig:axesratiodist}
\end{figure}

\begin{figure}
	\includegraphics[width=0.5\textwidth]{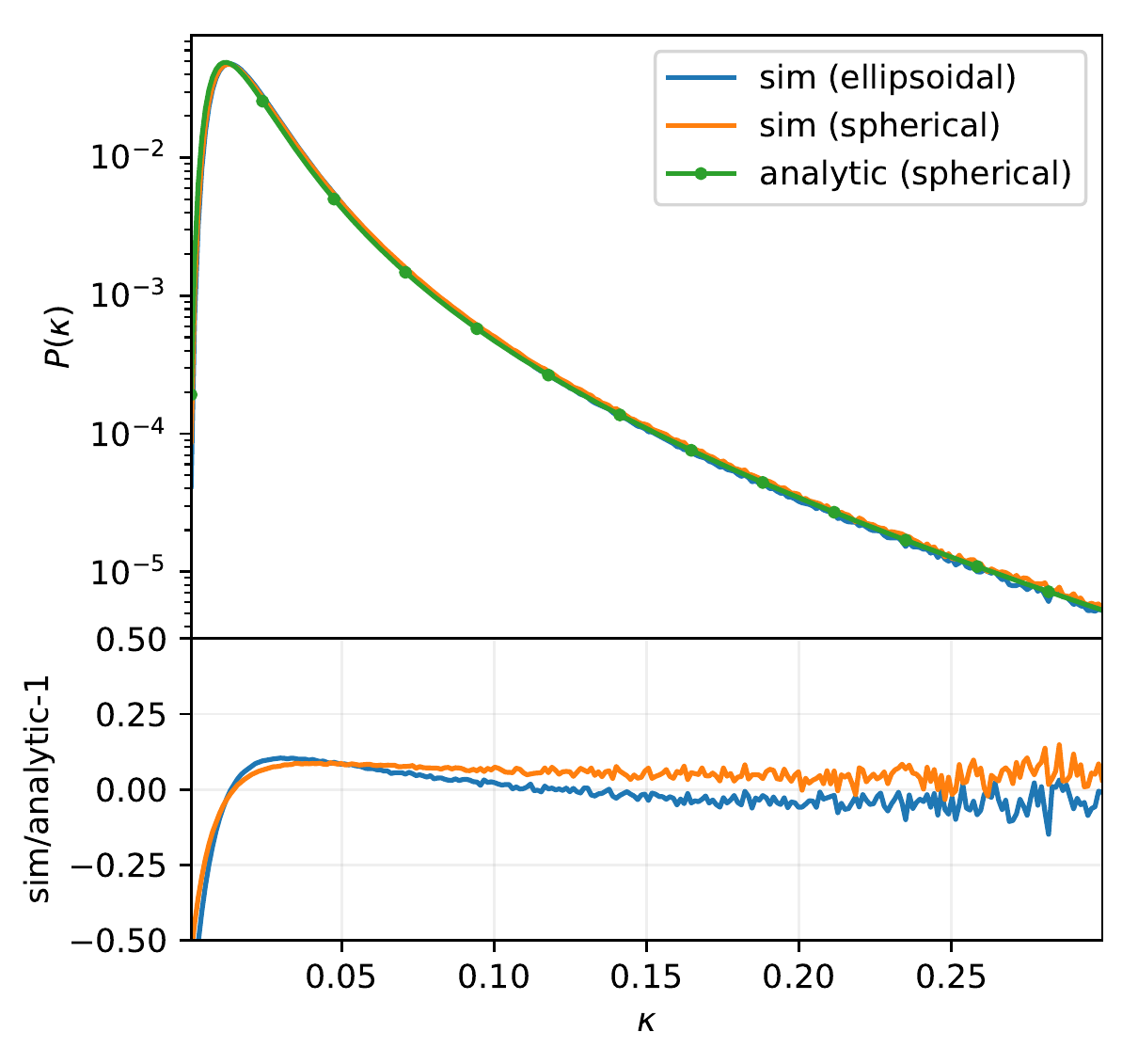}
	\caption{
	Here we explore the effect of including non-spherical halos
	on the PDF (here source redshift $z_{\text s}=1$).
	We see that the discrepancy between the complete ellipsoidal simplified simulations
	and the analytic result is relatively small, in particular in comparison to
	the discrepancy between our model and the MassiveNuS and T17 simulations.
	Note that in this exercise we have set a minimum mass cutoff at
	$\log M_\text{vir}/M_{\odot} = 12.5$ (i.e. higher than the fiducial choice of
	$\log M_\text{200m}/M_{\odot} = 11$).
	This is done purely in order to keep runtime of the simplified simulations reasonable,
	but since it is consistently implemented on the analytic side no issues arise from this
	choice.
	}
	\label{fig:triaxiality}
\end{figure}

The second technical assumption is the neglect of halo triaxiality. Our formalism could in principle
be adapted quite easily to allow for triaxiality, introducing more integrations over a
shape-distribution function and halo orientations.
However, these additional integrations, combined with the fact that the projected convergence
profiles would no longer be azimuthally symmetric, renders the computation substantially more
complex.
Thus, we explicitly test for the influence of triaxiality on the convergence PDF.
To this end, we measure the distribution of principal axes ratios in the MassiveNuS halo catalog.
This distribution is plotted in Fig.~\ref{fig:axesratiodist}.
We find that the relatively coarse binning in both mass and redshift is sufficient to capture the
variation of the shape distribution.
The extreme principal axes ratios for the lowest mass bin are likely driven by non-virialized
objects erroneously included in the halo catalog by the halo finder. However, this mass bin gives
only a negligible contribution to the PDF (cf. Sec.~\ref{subsec:MZcontributions}).

Then, we perform simplified simulations as described in more detail in Ref.~\cite{THS2019}.
In short, these simulations randomly populate maps with convergence profiles drawn from a given distribution
and measure the PDF in the end. By construction, the simplified simulations do not include the
clustering effect, but this is immaterial for the purposes of the test we want to perform here.
For the convergence profiles, we proceed as follows: First, for a halo of given mass and redshift,
we compute the NFW density profile. Then, we deform this spherically symmetric profile according to
ratios of principal axes drawn from the distribution described before. Finally, we perform a random
rotation of the resulting ellipsoid and then integrate the density along the line of sight to obtain
the convergence.
The result of this procedure is plotted in Fig.~\ref{fig:triaxiality}.
The green line with round markers is the analytic result assuming spherically symmetric profiles,
while the blue line is the described set of simplified simulations with ellipsoidal halos.
As a consistency test, we also run the simplified simulations with the principal axes distribution
set to a delta function at $b/a = c/a = 1$ (i.e. all halos are spherically symmetrical),
the result from this test is represented by the orange line.
First, we observe some discrepancies between the green and orange lines, which in principle we
should expect to coincide. However, there are two reasons why perfect agreement is not reached:
First, we remind the reader that our treatment of the quadratic pixels is not entirely correct;
and second, we expect some systematic errors in the numerical integration through the deformed NFW
profiles.
Thus, we consider this code test as passed.
Keeping this in mind, the discrepancies between the analytic result and the simplified simulations
incorporating triaxial halos are relatively small. Thus, triaxiality can certainly not account
for the discrepancies we observed between our model and the MassiveNuS as well as the T17
simulations. We conclude that more reliable modeling of small scale matter clustering is of much
greater importance than incorporating the small corrections from triaxiality.

\subsection{Substructure}
\label{subapp:substructure}

A final assumption is the lack of substructure in the density profiles.
Considering our results concerning triaxiality, it is reasonable to assume that, given the map
pixelization, the error incurred by ignoring substructure is relatively small as well.
Given the dramatic effect the Wiener filter in Sec.~\ref{sec:fisher} had in suppressing the
one-point PDF's sensitivity on the concentration model,
it is unlikely that parameter inference would require incorporation of substructure corrections.

\end{document}